\documentclass[a4paper,twocolumn,amssymb]{revtex4}
\usepackage{dcolumn}
\usepackage{bm}
\usepackage{graphicx,color}
\usepackage{epstopdf}
\usepackage{xspace}
\usepackage{amsmath}
\usepackage[english]{babel}
\usepackage{verbatim}
\usepackage{braket}
\usepackage{booktabs}
\usepackage{chemformula}
\let\ce \ch

\newcommand{\ali}{$\alpha_{i}$}

\newcommand{\alc}{$\alpha_{\mathrm{c}}$}
\newcommand{\li}{$\lambda_{i}$}

\newcommand{\TN}{$T_{\mathrm{N}}$}
\newcommand{\TS}{$T_{\mathrm{S}}$}

\newcommand{\fran}{Cu$_3$Bi(SeO$_3$)$_2$O$_2$Cl}
\newcommand{\Brfran}{Cu$_3$Bi(SeO$_3$)$_2$O$_2$Br}
\newcommand{\fr}{CBSCl}
\newcommand{\aimag}{$\alpha_{i,\mathrm{mag}}$}

\newcommand{\cp}{$c_{\mathrm{p}}$}

\newcommand{\cpmag}{$c_{\mathrm{p,mag}}$}

\begin{document}
	
	\title{Linear magnetoelastic coupling and magnetic phase diagrams of the buckled-kagom\'e antiferromagnet \ce{Cu3Bi(SeO3)2O2Cl}}
	\author{S. Spachmann$^{1,*}$}
	\author{P. Berdonosov$^{2,3}$}
	\author{M. Markina$^{2}$}
	\author{A. Vasiliev$^{2,3}$}
	\author{R. Klingeler$^{1}$}
	
	\affiliation{$^1$Kirchhoff Institute for Physics, Heidelberg University, D-69120 Heidelberg, Germany}
	\affiliation{$^2$Lomonosov Moscow State University, Moscow 119991, Russia}
	\affiliation{$^3$National University of Science and Technology "MISiS", Moscow 119049, Russia}
	\date{\today}

\begin{abstract}
\textbf{\abstractname.} 
Single crystals of \fran\ were investigated using high-resolution capacitance dilatometry in magnetic fields up to 15~T. Pronounced magnetoelastic coupling is found upon evolution of long-range antiferromagnetic order at \TN~$= 26.4(3)$~K. Gr\"{u}neisen analysis reveals moderate effects of uniaxial pressure on \TN, of $1.8(4)$~K/GPa, $-0.62(15)$~K/GPa and $0.33(10)$~K/GPa for $p \parallel a$, $b$, and $c$, respectively. Below 22~K, Gr\"{u}neisen scaling fails which implies the presence of competing interactions. The structural phase transition at \TS~$= 120.7(5)$~K is much more sensitive to uniaxial pressure than \TN, with strong effects of up to $27(3)$~K/GPa ($p \parallel c$). Magnetostriction and magnetization measurements reveal a linear magnetoelastic coupling for $B\parallel c$ below \TN, as well as a mixed phase behavior above the tricritical point around 0.4~T.
An analysis of the critical behavior in zero-field points to three-dimensional (3D) Ising-like magnetic ordering.
In addition, the magnetic phase diagrams for fields along the main crystalline axes are reported.
\end{abstract}

 \maketitle

\section{Introduction}
Frustrated magnetism has been a highly active research field in the past three decades~\cite{Lacroix2011}. Besides simple corner-sharing triangular lattice geometries, the Kagom\'{e}, pyrochlore, and hyperkagome lattices offer potential playgrounds of strongly geometrically frustrated magnetism~\cite{Ramirez1994, Mendels2016, Khomskii2014}.
Among them, the ideal $S = 1/2$ Kagom\'{e} Heisenberg antiferromagnet is the most prominent realization of a system in which macroscopic ground state degeneracies, i.e., a (quantum) spin liquid state, are expected to prevent the evolution of any long-range order~\cite{Savary2016}.
Other phenomena of frustrated magnetism include fractionalized magnetization plateaus, chiral and helical spin arrangements, as well as spin glass, spin nematic, and spin ice behaviors~\cite{Lacroix2011}.
A number of geometrically frustrated systems such as \ce{FeTe2O5Cl}~\cite{Pregelj2013}, \ce{PbCu3TeO7}~\cite{Yoo2018}, and \ce{Ni3V2O8}~\cite{Cabrera2009, Zhang2013} were found to additionally exhibit multiferroic behavior. From a technological perspective multiferroics -- materials combining more than one ferroic property such as ferromagnetism, ferroelectricity and ferroelasticity~\cite{Spaldin2005, Fiebig2016} -- are especially sought-after for the control of magnetism via electric fields, with the promise of substantially lower energy consumption than manipulating magnetic states via magnetic fields~\cite{Manipatruni2018}.
Potential applications range from ultra-low power logic-memory~\cite{Manipatruni2019} to radio- and high-frequency devices, including electric field-tunable radio-frequency/microwave signal processing, magnetic field sensors, magnetoelectric random access memory (MERAM)~\cite{Bibes2008} and voltage-tunable magnetoresistance~\cite{Liu2014, Spaldin2019}.

\fran, eponym of the francisite family~\cite{Becker2005, Zakharov2014, Zakharov2016, Klimin2017, Markina2017, Markina2021}, crystallizes in a special buckled realization of the Kagom\'{e} lattice~\cite{Pring1990} and exhibits multiferroic properties at low temperatures~\cite{Wu2017, Constable2017}.
Cu$^{2+}$ ions ($3d^{9}, S$~=~1/2) situated at two different crystal sites, Cu1 and Cu2, are the magnetic centers forming the Kagom\'{e} lattice in the $ab$ plane.
Both Cu1 and Cu2 ions are found in a square planar coordination with Cu-O bond lengths of 1.933~\r{A} to 1.978~\r{A}. These plaquettes around Cu1 and Cu2 sites are non-parallel with respect to each other.
Along the $c$ axis Cu ions are connected by long Bi--O bonds with a bond length of about 2.8~\r{A}.
Below \TN~$\approx 26$~K \fran\ (in short: \fr) develops an A-type antiferromagnetic order along the $c$ axis of spins aligned ferromagnetically (FM) in the $ab$ plane. Cu2 spins are aligned strictly (anti)parallel with the $c$ axis whereas Cu1 spins are canted towards the $b$ axis by 59(4)$^{\circ}$~\cite{Constable2017}.
This peculiar magnetic order arises from the competition of large FM nearest neighbor exchange interactions ($J_1$ and $J_1'$, see Fig.~S2 [\onlinecite{Rousochatzakis2015}]) on the order of $-70$~K to $-80$~K and AFM next-nearest neighbor interactions ($J_2$) on the order of 60~K, in conjunction with small exchange couplings along the $c$ direction ($J_{\perp,1}$ and $J_{\perp,2}$, $|J_{\perp,i}|\leq 2$~K), Dzyaloshinskii-Moriya interactions and a symmetric anisotropic exchange component~\cite{Rousochatzakis2015, Nikolaev2016, Constable2017}.
Furthermore, a linear magnetoelectric coupling has been observed below \TN~\cite{Constable2017, Wu2017}.
At higher temperatures around 120~K a structural phase transition from an orthorhombic $Pmmn$ to an orthorhombic and nonpolar, possibly antiferroelectric (AFE), $Pcmn$ space group occurs upon cooling~\cite{Constable2017, Gnezdilov2017, Milesi-Brault2020}.

In this paper we investigate the interplay of the lattice and spin degrees of freedom in \fr\ with high-resolution thermal expansion and magnetostriction as well as magnetization measurements. From these measurements we construct the complete magnetic phase diagram for the first time, and quantify the magnetoelastic coupling by the effects of uniaxial pressure on the phase boundaries. Furthermore, magnetostriction and magnetization measurements reveal a linear magnetoelastic coupling below \TN. Finally, a critical scaling analysis suggests that three-dimensional Ising-like magnetic ordering occurs at \TN, whereas magnetic correlations above \TN\ are constrained to the $ab$ plane.

\section{Results and Discussion}
\subsection{Thermal Expansion}
\begin{figure*}[t]
	\center{\includegraphics [width=0.95\columnwidth,clip]{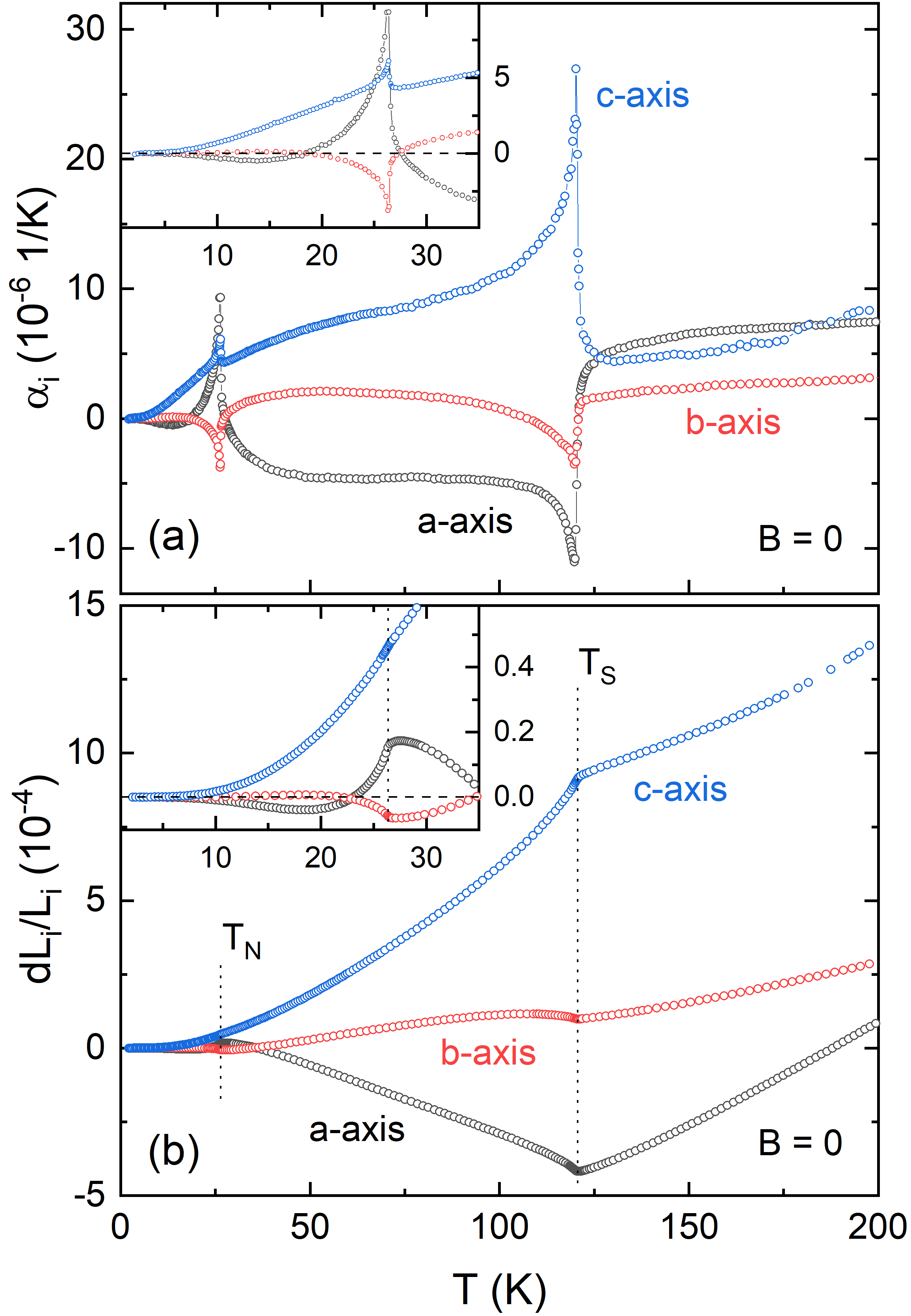}}
	\caption[] {\label{alpha_0T}\textbf{Zero-field thermal expansion:} (a) Thermal expansion coefficients \ali\ and (b) relative length changes $dL_i/L_i$ in zero-field for the crystallographic $a$, $b$, and $c$ axis of \fr. Insets show a magnification of the low temperature regime around \TN\ with zero y-values indicated by horizontal dashed lines. Vertical dotted lines in (b) mark \TN\ and \TS.}
\end{figure*}

\textbf{Zero-field:} Thermal expansion measurements in zero field show pronounced anomalies at \TN~$= 26.4(3)$~K and \TS~$= 120.7(5)$~K (Fig.~\ref{alpha_0T}).
At \TN\ these anomalies provide evidence for the presence of significant magnetoelastic coupling in \fr.
In the thermal expansion coefficients \ali\ the anomalies have $\lambda$-like shapes, signaling continuous phase transitions at both \TN\ and \TS. Notably, the transition at \TS\ shows a jump superimposing the $\lambda$-like behavior for the $a$ and $c$ axis, but not for the $b$ axis.
The length changes observed in \fr\ (Fig.~\ref{alpha_0T}(b)) are highly anisotropic. While the $a$ axis shrinks above \TN\ upon warming up to \TS, the $c$ axis strongly expands in this temperature regime. Expansion along the $b$ axis, in contrast, is much smaller than for the other two axes.
The volume expansion is positive in the whole measured temperature regime (see supplement, Fig.~S3).

\textbf{Gr\"{u}neisen analysis:} The magnetoelastic coupling in \fr\ can be quantified by means of the uniaxial pressure dependence of \TN. This pressure dependence may be derived from the magnetic contributions to thermal expansion and specific heat using the Ehrenfest relation which is valid for continuous phase transitions~\cite{Barron-White1999, Klingeler2005}:
\begin{equation}\label{eq:gruen}
    \frac{{\partial}T_{\mathrm{N}}}{{\partial}p_{i}} = T_{\mathrm{N}}V_{\mathrm{m}} \gamma_{\mathrm{(mag)}} =  T_{\mathrm{N}}V_{\mathrm{m}} \frac{\alpha_{i(,\mathrm{mag})}}{c_{\mathrm{p(,mag)}}}.
\end{equation}
$\gamma_{\mathrm{(mag)}}$ is the (magnetic) Gr\"{u}neisen ratio and $V_{\mathrm{m}}$ the molar volume, which for \fr\ is $V_{\mathrm{m}} = 1.333\cdot 10^{-4}$~m${^3}$/mol~\cite{Millet2001}. The magnetic contributions to the thermal expansion, \aimag, and specific heat, \cpmag, in Fig.~\ref{Gruen} were obtained by subtracting a phononic background (see supplement, Fig.~S4) estimated by a sum of Debye and Einstein functions (for more details see supplement).
In order to assess the validity of a constant $\gamma_{\mathrm{(mag)}}$ in Eq.~\eqref{eq:gruen}, in Fig.~\ref{Gruen} the ordinates are scaled for an optimal overlap of \aimag\ and \cpmag\ around \TN. Above \TN\ and down to about 22~K (24~K for $c$) the data scale well within the experimental errors which implies a single dominating energy scale in this temperature regime~\cite{Barron-White1999}. Below 22~K, however, a difference in the behavior of \aimag\ and \cpmag\ is clearly visible for all axes as well as the volume. This difference indicates competing interactions arising below \TN. A discussion of possible sources for this behavior is given in the supplemental material.
\begin{figure*}[t]
	\center{\includegraphics [width=0.95\columnwidth,clip]{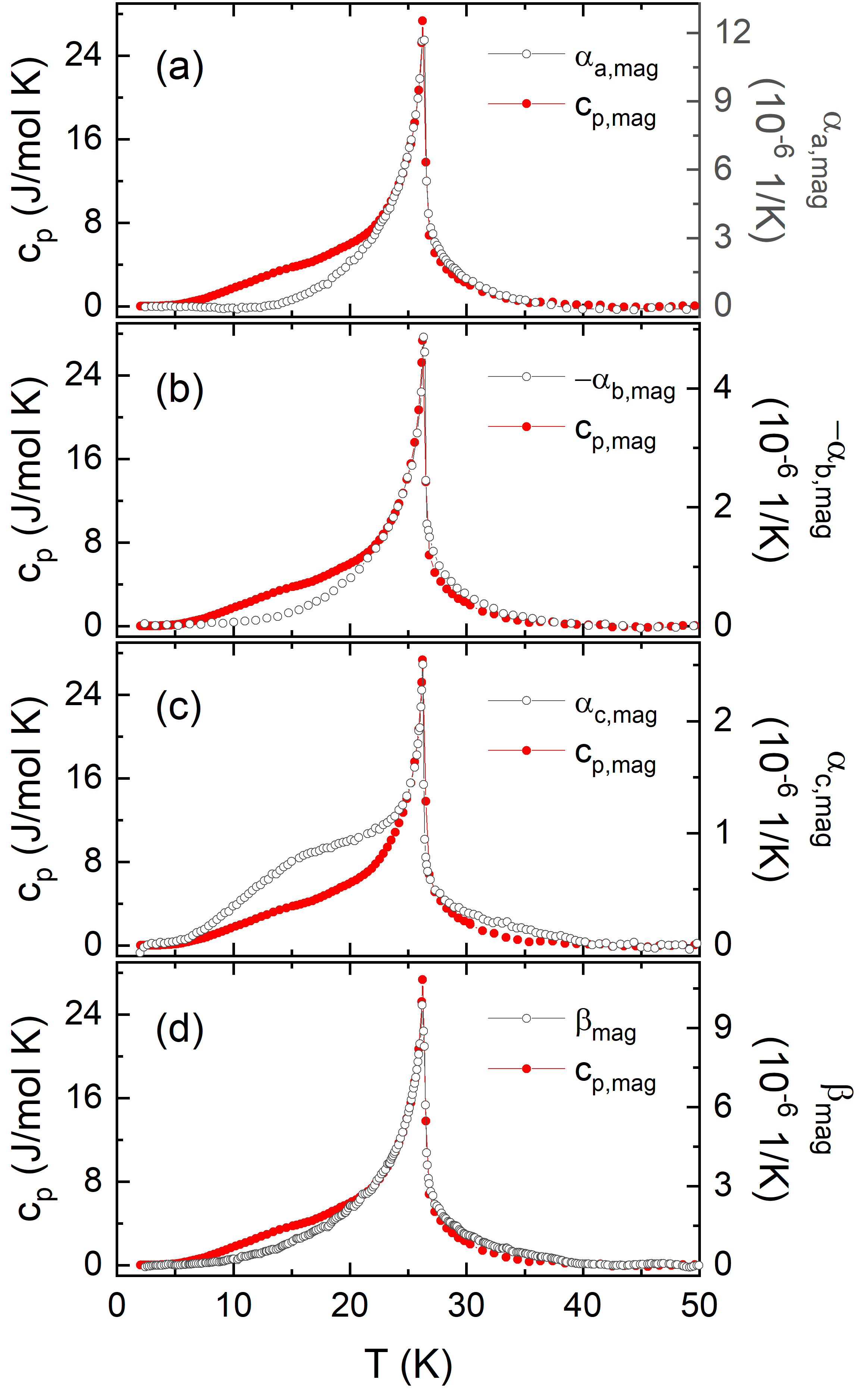}}
	\caption[] {\label{Gruen}\textbf{Gr\"{u}neisen Scaling:} Comparison of magnetic contributions to the specific heat (red circles, left axes) and thermal expansion (empty black circles, right axes) along the (a) $a$ axis, (b) $b$ axis, (c) $c$ axis, and (d) for the volume expansion.}
\end{figure*}

The magnetic Gr\"{u}neisen ratio in Eq.~\eqref{eq:gruen} is obtained from the ratio of the scaled ordinates in Fig.~\ref{Gruen}. This results in $\gamma_{\mathrm{a,mag}} = 5.0\cdot 10^{-7}$~mol/J, $\gamma_{\mathrm{b,mag}} = -1.8\cdot 10^{-7}$~mol/J, and $\gamma_{\mathrm{c,mag}} = 9.3\cdot 10^{-8}$~mol/J, which yields uniaxial pressure dependencies at \TN\ of 1.8(4)~K/GPa ($p\parallel a$), $-0.62(15)$~K/GPa ($p\parallel b$) and 0.33(10)~K/GPa ($p\parallel c$). Long-range antiferromagnetic (AFM) order is thus strengthened by pressure $p\parallel a$ and $c$ whereas $p\parallel b$ influences the competition of interactions such that long-range order is suppressed. From a qualitative point of view, increase of \TN\ upon in-plane pressure which likely increases the lattice along the $c$ direction suggests that the perpendicular exchange couplings ($J_{\perp}$) are not crucially driving long-range AFM order. We also speculate that uniaxial pressure $p\parallel a$ ($p\parallel b$) changes the Cu1-O-Cu1 bond angle further towards (away from) 90$^\circ$, thereby increasing (decreasing) ferromagnetic exchange $J_1$ while $J_1'$ becomes smaller for $p\parallel a$. Our results, i.e., ${\partial}T_{\mathrm{N}}/{\partial}p_{a} > 0$ and ${\partial}T_{\mathrm{N}}/{\partial}p_{b} < 0$ hence indicate a crucial role of $J_1$ for stabilizing A-type AFM order.

The analysis of the structural phase transition yields much larger values for the uniaxial pressure dependencies ${\partial}T_{\mathrm{S}}/{\partial}p_i$ in comparison with ${\partial}T_{\mathrm{N}}/{\partial}p_i$. In-plane pressure strongly suppresses the structural phase transition (${\partial}T_{\mathrm{S}}/{\partial}p_a = -19(6)$~K/GPa, ${\partial}T_{\mathrm{S}}/{\partial}p_b = -5.7(1.1)$~K/GPa) whereas pressure applied out-of-plane enhances \TS\ even more strongly (${\partial}T_{\mathrm{S}}/{\partial}p_c = 27~K$/GPa). The Gr\"{u}neisen ratios corresponding to these values are derived in a different manner than for \TN, by area-conserving -- and for \cp\ entropy-conserving -- interpolations of the jump heights of the respective quantities at \TS\ (see supplement, Fig.~S5).
Table~\ref{tab:Gruen} summarizes the pressure dependence of the transition temperatures in \fr.

\renewcommand{\arraystretch}{1.1}
\begin{table}[ht]
    \centering
    \caption{\label{tab:Gruen} Pressure dependencies of the transition temperatures calculated from the magnetic Gr\"{u}neisen ratios $\gamma_{i,\mathrm{mag}}$ at \TN, as well as from the jumps $\Delta\alpha_i$ and ${\Delta}c_{\mathrm{p}}$ at \TS, for the three main crystallographic axes. The hydrostatic pressure dependence $dT_{\mathrm{S}}/dp$ is calculated as the sum of all three uniaxial ones and denoted as "volume".}
    \begin{tabular}{l|cc|cc}
    \hline \hline
         & $\gamma_{i,\mathrm{mag}}$ & ${\partial}T_{\mathrm{N}}/{\partial}p_i$ & $\gamma_i$ & ${\partial}T_{\mathrm{S}}/{\partial}p_i$ \\
          & (10$^{-7}$ mol/J) & (K/GPa) & (10$^{-7}$ mol/J) & (K/GPa)\\ \hline
        $a$ axis & 5.0 & 1.8$\pm$0.4 &  --11.9 & --19$\pm$6  \\ 
        $b$ axis & --1.8 & --0.62$\pm$0.15 &  --3.5 & --5.7$\pm$1.1  \\
        $c$ axis & 0.93 & 0.33$\pm$0.10 & 16.9 & 27$\pm$3 \\
        volume & 4.0 & 1.4$\pm$0.3 & -- & 2.3$\pm$1.0 \\\hline
    \end{tabular}
\end{table}
\renewcommand{\arraystretch}{1}

We find that both \TN\ and \TS\ are enhanced by hydrostatic pressure. Previous studies investigated the effects of chemical doping on the physical properties of \fr\ by substituting the lone-pair Bi site by \ce{Y}~\cite{Zakharov2014} or lanthanide elements~\cite{Markina2017, Zakharov2016}, exchanging the halide ion for \ce{Br} or \ce{I}~\cite{Millet2001}, or \ce{Te} doping the \ce{Se} site~\cite{Wu2017}. Hydrostatic pressure applied to \fr\ was shown to enhance \TN\ by about 1~K/GPa~\cite{Wu2017} which is in line with our results. For the pressure dependence of \TS\ no literature data are available. The structural phase transition is absent in the compounds where \ce{Cl} is exchanged for \ce{Br} and \ce{I} due to their larger ionic radii, and was not reported for any of the other doped versions of francisite.

\begin{figure*}[t]
	\center{\includegraphics[width=0.95\columnwidth,clip]{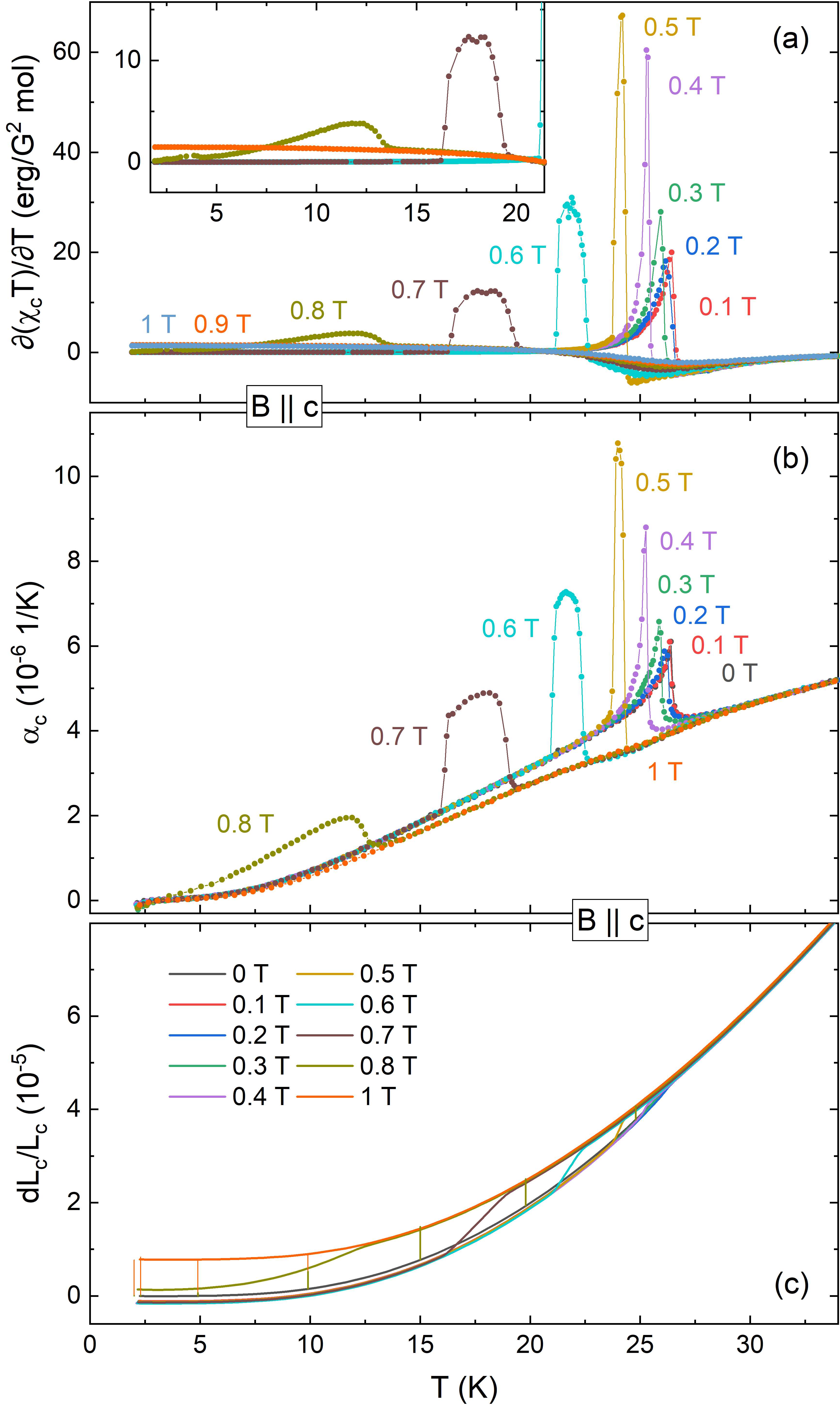}}
	\caption[] {\label{TE-and-M_c-axis} \textbf{Fisher's specific heat and thermal expansion up to $\mathbf{B = 1}$~T:} (a) Fisher's specific heat calculated from the static magnetic susceptibility, (b) thermal expansion coefficient \alc, and (c) relative length changes, for 0~T~$\leq B\parallel c \leq$~1~T. The inset in (a) presents a magnification of the low temperature region. Vertical bars in (c) mark the relative length changes ${\Delta}L(B)$ from 0~T to 0.8~T (olive green) and 1~T (orange), respectively, obtained from magnetostriction measurements.}
\end{figure*}

\textbf{Thermal Expansion for $\mathbf{B > 0, B\parallel c}$:} The $c$ axis is the easy-axis in \fr\ and shows a metamagnetic transition in magnetic fields $B\parallel c$, rendering it of special interest. 
Application of a magnetic field $B\parallel c$ suppresses the N\'{e}el transition to lower temperatures (Fig.~\ref{TE-and-M_c-axis}(b) and (c)). Up to a field of 0.5~T an increase in the thermal expansion coefficient signals field-induced enhancement of the pressure dependence of the entropy changes ($\alpha_i \propto {\partial}S/{\partial}p_i$). At even higher fields the character of the phase transition changes from a continuous to a discontinuous transition. This change is evidenced by a broadening of the $\lambda$-like feature to a plateau-like behavior between two jumps (up and down) in \alc\ as the temperature is increased. A magnetic field of 1~T fully suppresses AFM order while a ferromagnetic phase appears. Notably, the exact same behavior described for \alc\ at low fields is also visible in the Fisher specific heat~\cite{Fisher1962}, ${\partial}(\chi_{c}T)/{\partial}T$, a derivative of the static magnetic susceptibility $\chi_c$ (Fig.~\ref{TE-and-M_c-axis}(a)). This behavior and especially the intermediate phase emerging in an applied field will be discussed in more detail below.

\subsection{Magnetostriction for $B\parallel c$}
\begin{figure*}[t]
	\center{\includegraphics [width=1.5\columnwidth,clip]{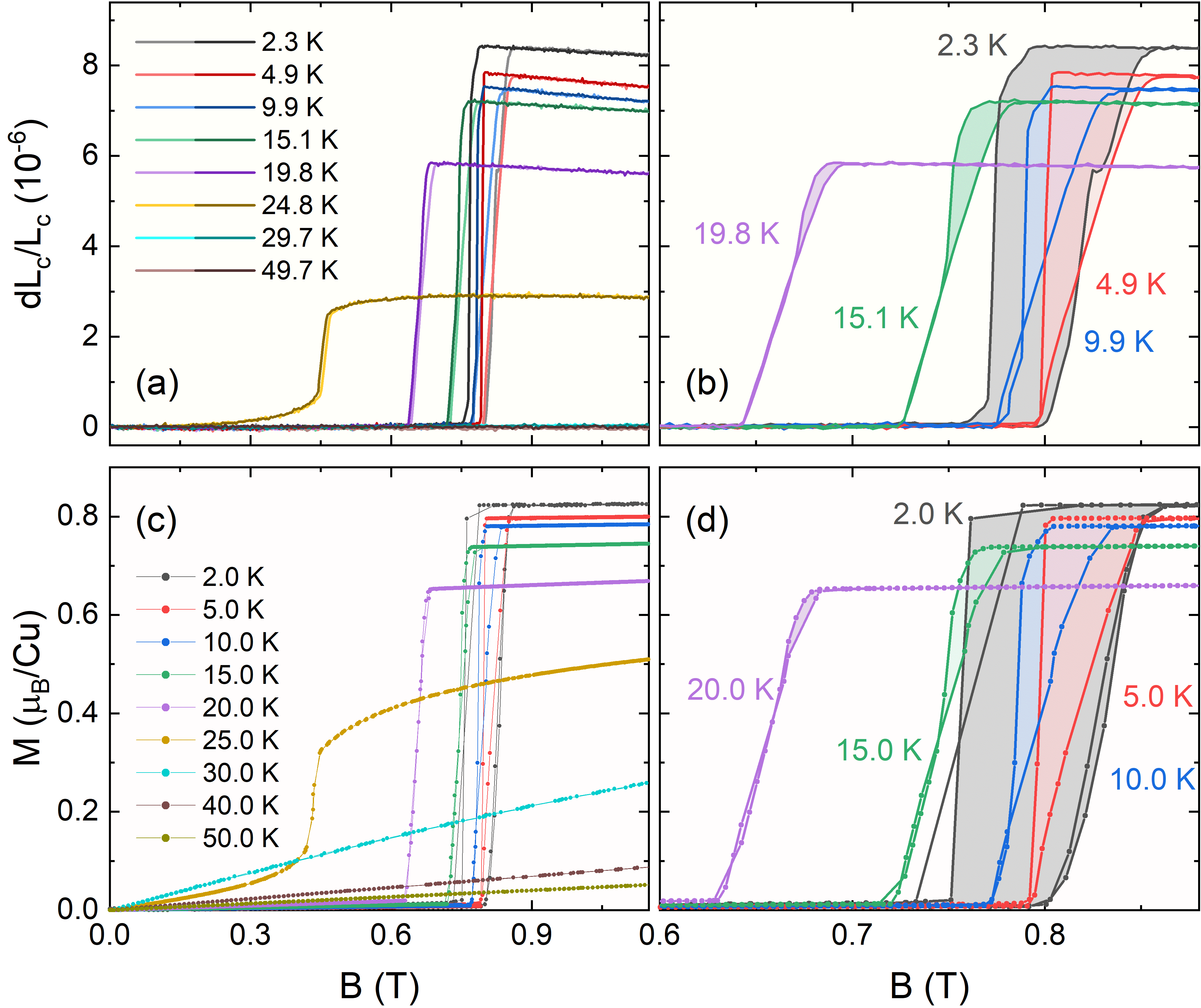}}
	\caption[] {\label{MS-and-M_c-axis}\textbf{Linear magnetoelastic coupling: Magnetostriction and magnetization.} Comparison of relative length changes $dL_{\mathrm{c}}(B)/L_{\mathrm{c}}(0)$ (a, b) and isothermal magnetization (c, d) for $B\parallel c$. Note that (b) and (d) show a magnification of the transition region and only data for selected temperatures. Hysteresis is marked by colored areas. Magnetostriction down-sweeps in (b) are shifted by 6~mT to correct for the remanent field of the magnet.}
\end{figure*}

Application of magnetic fields $B\parallel c$ drives the system into a ferromagnetic (FM) phase as indicated by sharp jumps in the magnetization (Fig.~\ref{MS-and-M_c-axis}(c, d)). The metamagnetic transition is characterized by sharp jumps in the macroscopic length as seen in the magnetostriction measurements along the $c$ axis (Fig.~\ref{MS-and-M_c-axis}(a, b)). In both magnetization and length changes, a broad hysteresis region becomes visible upon ramping down the applied field at 2~K (0.86~T to 0.77~T), indicating the first-order nature of the transition. The FM state features the magnetization of $M_c = 0.87~\mu_{\mathrm{B}}/$\ce{Cu} thereby indicating that the spins are not fully aligned but presumably canted.
At 10~K and above a linear increase is visible in $dL_c(B)$ between a regime where $dL_c \approx 0$ at low fields and a constant value above the phase transition. The hysteresis region shrinks as the temperature is increased towards \TN\ and does not extend over the whole regime of linear increase.

\textbf{Mixed-phase for $\mathbf{B > 0.4~T}$:} The linear increase and hysteresis observed in $dL_c(B, T < T_{\mathrm{crit}})$ points to a mixed phase between the low-field AFM ordered phase and the field-induced FM state. In this mixed intermediate phase both AFM and FM domains are present and the ratio FM:AFM linearly increases with the applied magnetic field, as the demagnetizing field is overcome and AFM regions align with the field.
Such mixed-phase behavior is common for metamagnets and has also been observed in the brother compound \Brfran~\cite{Pregelj2015}. There, the mixed-phase was found to exhibit broadband absorption with excitations extending over at least ten decades of frequency~\cite{Pregelj2015}. A similar behavior is naturally also expected in \fran.

\begin{figure*}[t]
	\center{\includegraphics [width=1.2\columnwidth,clip]{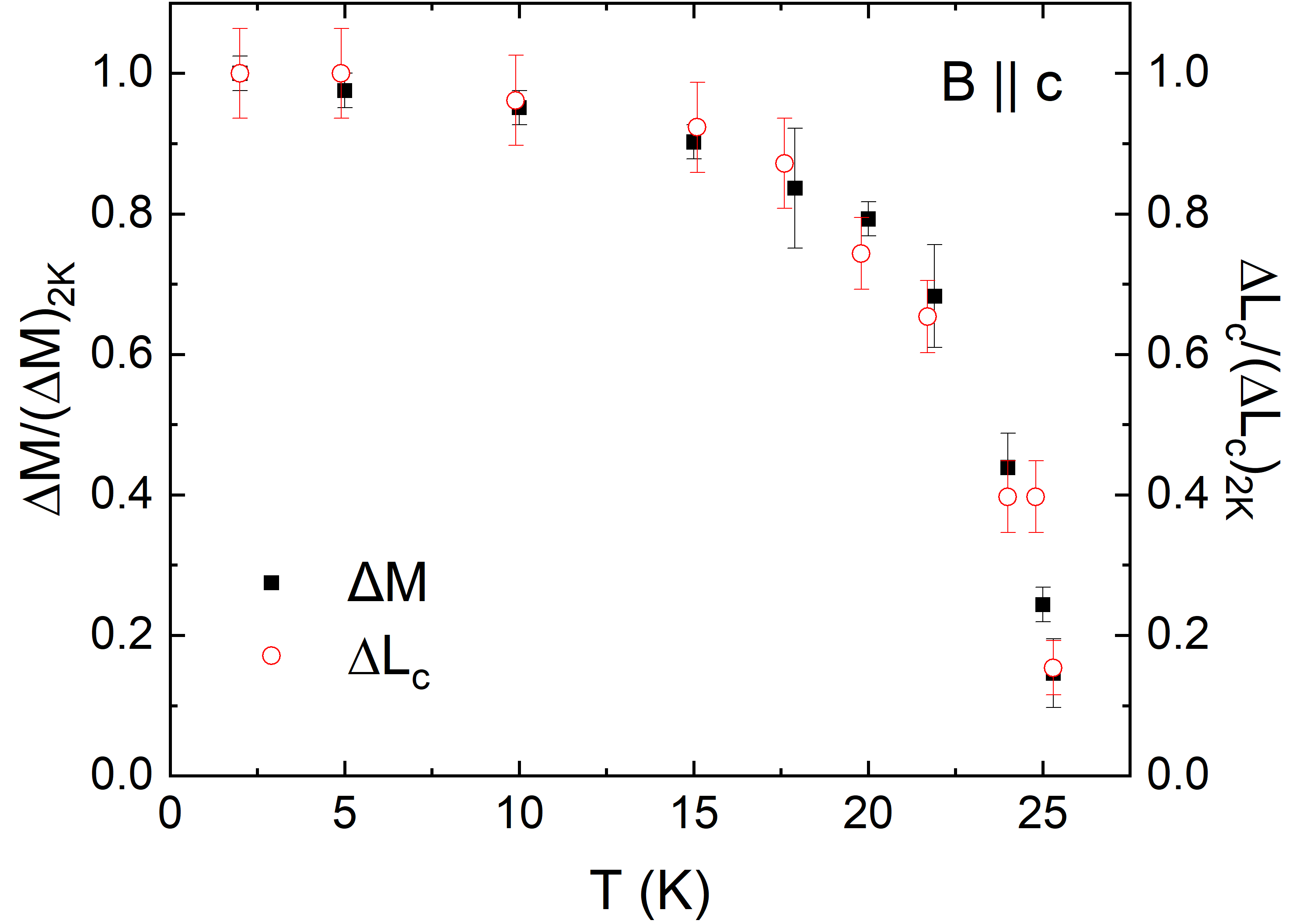}}
	\caption[] {\label{c-axis_dM-vs-dLL}\textbf{Linear magnetoelastic coupling: Jump sizes.} Comparison of the jumps in magnetization (left ordinate) and relative length changes (right ordinate) at the metamagnetic transition normalized by their value at 2~K.}
\end{figure*}

\textbf{Linear magnetoelastic coupling:} On top of the mixed-phase behavior, a remarkable direct proportionality between the magnetization $M_c$ and the relative length changes $dL_c(B)$, i.e., a linear magnetoelastic coupling $dL_c(M_c) = 9\cdot  10^{-6} M_c$~[$\mu_{\mathrm{B}}$/Cu], is observed (Fig.~\ref{MS-and-M_c-axis}) at $T < T_{\mathrm{crit}}$. A comparison of the jump heights at the metamagnetic transition, normalized to the jump heights at 2~K, illustrates this observation even more clearly (Fig.~\ref{c-axis_dM-vs-dLL}). The temperature evolution of both $\Delta M$ and $\Delta L_c$ are proportional to each other and obeys an order parameter-like behavior.
The phenomenon of linear magnetoelastic coupling was investigated extensively in the late 1950s and early 1960s, both in ferro- and antiferromagnetic materials.
According to Zvezdin et al.~\cite{Zvezdin1985}, linear magnetostriction was first observed in \ce{CoF2} by Borovik-Romanov and co-workers~\cite{Borovik-Romanov1963}.
Among many others, antiferromagnetic examples include hematite (${\alpha}$-Fe$_{2}$O$_{3}$)~\cite{Scott1966}, dysprosium orthoferrite~\cite{Zvezdin1985} (\ce{DyFeO3}), and more recently the multiferroic \ce{TbMnO3}~\cite{Aliouane2006}.
In all of these compounds linear magnetostriction is closely related to the presence of antiferromagnetic domains.
The magnetic point group of \fr, $mm'm$, is not among the point groups in which a linear magnetostriction is expected to be dominant~\cite{Birss1963}, which does not mean, however, that it may not, as observed, be dominant.

In addition to the linear magnetoelastic coupling, it was argued that the point group $mm'm$ is also compatible with a linear magnetoelectric coupling~\cite{Constable2017}, and indeed, a linear magnetoelectric coupling was shown experimentally~\cite{Wu2017}.

\textbf{In-plane magnetic correlations far above $\mathbf{T_{\mathrm{N}}}$:} Thermal expansion measurements at 15~T reveal a large temperature regime above \TN\ where significant in-plane magnetostriction is present (see supplement, Fig.~S7). Notably, magnetostriction along the $c$ axis is negligible above \TN, whereas for the $a$ and $b$ axis there are pronounced magnetic field effects up to above 80~K. For the $a$ axis the length changes are about one magnitude larger than for the $b$ axis. Qualitatively, application of a magnetic field acts in the same way as the evolution of long-range order below \TN: the $a$ axis shrinks whereas the $b$ axis elongates.
The strong difference between magnetostriction in the $ab$ plane and along the $c$ axis evidences strong in-plane magnetic correlations above \TN, in line with the large and competing FM and AFM exchange couplings on the order of 60~K to 70~K and the small inter-plane couplings on the order of up to 2~K~\cite{Rousochatzakis2015,Nikolaev2016}.
Negligible magnetostriction along the out-of-plane direction, $\lambda_c = -{\partial}M_c/{\partial}p_c \sim {\partial}J/{\partial}p_c$ signals that the exchange interactions $J$ are nearly independent of pressure $p\parallel c$.

\subsection{Phase Diagrams}
The low-temperature magnetic phase diagrams for \fr\ as measured by thermal expansion ($c$ axis) and magnetization (all axes) are presented in Fig.~\ref{PD}. Corresponding magnetization measurements are shown in the supplement (Fig.~S9, S10, and S11). The saturation field of $B_{\mathrm{sat, a}} = 20.7(5)$~T is extrapolated from the data for $B\parallel a$, whereas $B_{\mathrm{sat, b}} = 7.1(2)$~T and $B_{\mathrm{sat, c}} = 0.86(2)$~T were obtained by measurements at 2~K for $B\parallel b$ and $B\parallel c$, respectively.
For the $c$ axis, a change in behavior from a continuous phase transition at zero- and low fields to a discontinuous transition above 0.4~T signals the presence of a tricritical point around 0.4~T. Moreover, a large magnetic field hysteresis at the metamagnetic transition is visible at low temperatures, which decreases as the temperature is increased.
\begin{figure*}[htbp]
	\center{\includegraphics [width=2\columnwidth,clip]{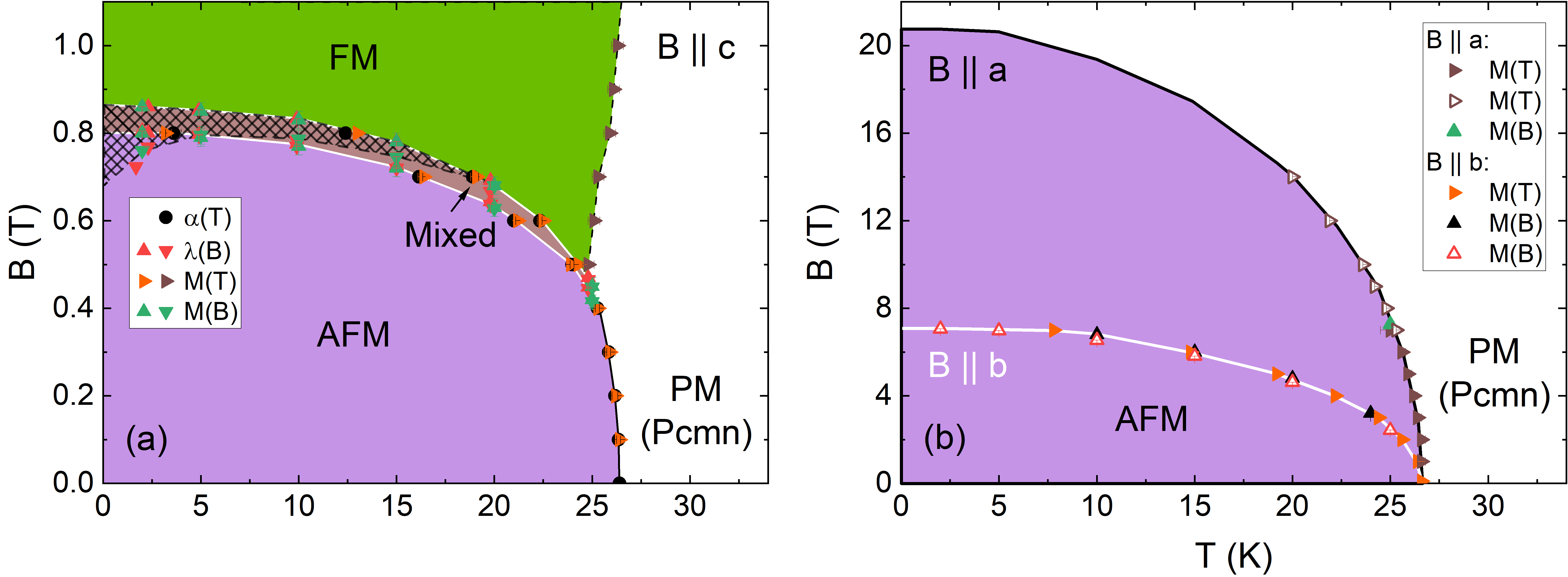}}
	\caption[] {\label{PD}\textbf{Magnetic Phase Diagrams:} (a) Low-temperature magnetic phase diagram for $B\parallel c$. The hatched area displays the hysteresis region visible in isothermal magnetization and magnetostriction curves. Where hysteresis and the mixed phase overlap, white lines indicate phase boundaries obtained from up-sweep data. (b) Low-temperature magnetic phase diagram for $B\parallel a$ (black boundary) and $B\parallel b$ (white boundary) constructed from magnetization measurements. The phase boundary \TN ($B\parallel a>14$~T) is obtained by scaling \TN ($B\parallel b$).}
\end{figure*}

\textbf{Quantitative analysis of the phase boundaries for $\mathbf{B}\parallel \mathbf{c}$:} By measuring the jump-like discontinuities in the magnetization and length at the metamagnetic phase boundary $B_{\mathrm{crit}}(T)$, the uniaxial pressure dependencies of the critical field and the critical temperature can be derived from the Clausius-Clapeyron equation and related thermodynamic equations~\cite{Barron-White1999}
\begin{align}\label{eq:fran_Clapeyron}
	\left(\frac{{\partial}T_{\mathrm{crit}}}{{\partial}p_c}\right)_B &= V_{\mathrm{m}}\frac{\frac{{\Delta}L_c}{L_c}}{\Delta S} \\[5pt]
	\left(\frac{{\partial}T_{\mathrm{crit}}}{{\partial}B_c}\right)_p &= -\frac{\Delta m_c}{\Delta S} = -\frac{\Delta (M_c\cdot V)}{\Delta S}\\[5pt]
	\left(\frac{{\partial}B_{\mathrm{crit}}}{{\partial}p_c}\right)_T &= V_{\mathrm{m}}\frac{\frac{{\Delta}L_c}{L_c}}{\Delta m_c}
\end{align}
where ${\Delta}L_c/L_c$ and ${\Delta}m_c$ are the observed jumps in the length and the magnetic moment. ${\partial}T_{\mathrm{crit}}/{\partial}B_c$ is calculated from two polynomial fits in different temperature regimes to the phase boundary $B_{\mathrm{crit}}(T)$ such that the changes in entropy, ${\Delta}S$, can be calculated.
The pressure dependence of the critical field $B_{\mathrm{crit}}$ is small and positive, on the order of 80(6)~mT/GPa at all temperatures. Derived changes of the transition temperature $T_{\mathrm{crit}}(B)$ at the metamagentic (AFM to FM) phase transition under pressure are large at low temperatures (${\partial}T_{\mathrm{crit}}/{\partial}p_c = 34(16)$~K/GPa at 2~K) and decrease strongly as the temperature is increased  (${\partial}T_{\mathrm{crit}}/{\partial}p_c = 1.1(5)$~K/GPa at 24~K).
Correspondingly, the entropy related to the phase transition increases from 30~mJ/(mol K) at 2~K to 360(130)~mJ/(mol K) at 24~K.
This analysis shows that uniaxial pressure $p\parallel c$ changes the exchange couplings in such a way that the AFM phase is stabilized with respect to both the FM and the paramagnetic (PM) phase. 
An overview of calculated values is given in the supplement in Tab.~S2.

\subsection{Critical Scaling}
In the vicinity of a critical point the specific heat is expected to behave as
\begin{align*}
    c_{\mathrm{p}} &= A\cdot t^{-\alpha}+ B &T > T_{\mathrm{crit}} \\
    c_{\mathrm{p}} &= A'\cdot |t|^{-\alpha'}+ B' &T < T_{\mathrm{crit}}
\end{align*}
where $t = T/T_{\mathrm{crit}} - 1$ is the reduced temperature~\cite{Kadanoff1967}.
As seen in the Gr\"{u}neisen scaling in Fig.~\ref{Gruen}, the magnetic contributions to the specific heat and the thermal expansion around \TN\ can be scaled to each other.
Therefore, we can safely assume that a critical scaling of \aimag\ close to \TN, i.e., above the shoulder-like feature at low temperatures ($t < -0.2$ for $i = a, b$ and $t < -0.06$ for $i = c$), with the expression from Eq.~\eqref{eq:Scaling} is allowed.
Although we use the magnetic contributions, \cpmag\ and \aimag, for the fitting, there is an uncertainty on the phonon background correction.
Therefore, and to account for possible further contributions on top of the critical behavior, we initially used the canonical expression (similar to the one in Ref.~\cite{Kornblit1973})
\begin{equation}\label{eq:Scaling}
    c_{\mathrm{p}} = \frac{A^\pm}{\alpha^\pm}|t|^{-\alpha^\pm}(1+E^{\pm}|t|^{0.5})+B+D^{\pm}t
\end{equation}
where ''$+$'' (''$-$'') denotes fitting parameters for $T > T_{\mathrm{crit}}$ ($T < T_{\mathrm{crit}}$).
It turned out, however, that the data can be fitted very well even when setting $B = 0$ and $D^\pm = 0$.

Depending on the critical exponent $\alpha^\pm$ and the ratio of the amplitudes, $A^{+}/A^{-}$, the critical behavior can be categorized by one of many universality classes. 
The most well-known universality classes for a $d$-dimensional lattice and an order parameter of dimensionality $D$ are the 3D Heisenberg model ($d = 3$, $D = 3$), the 3D XY model ($d = 3$, $D = 2$), the 3D Ising model ($d = 3$, $D = 1$) and the 2D Ising model ($d = 2$, $D = 1$).
Calculations for these different models predict $\alpha \approx -0.12$, $A^{+}/A^{-} \approx 1.5$ (3D Heisenberg), $\alpha \approx -0.01$, $A^{+}/A^{-} \approx 1$ (3D XY), $\alpha \approx 0.11$, $A^{+}/A^{-} \approx 0.5$ (3D Ising)~\cite{Pelissetto2002} , as well as $\alpha = 0$ (2D Ising)~\cite{LeGuillou1980}.

\begin{figure*}[t]
	\center{\includegraphics [width=2\columnwidth,clip]{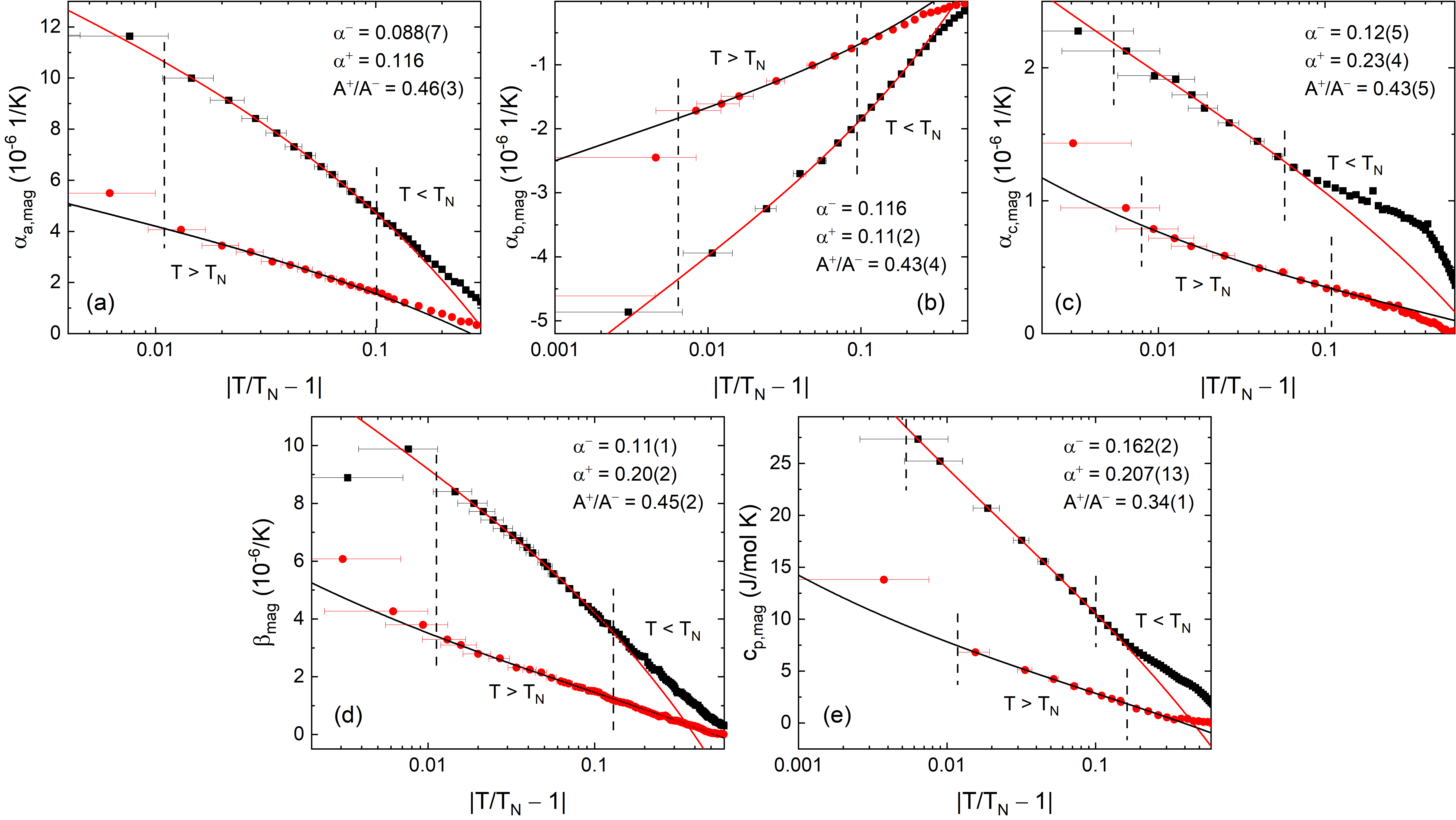}}
	\caption[] {\label{T_N_CritExp}\textbf{Critical scaling analysis near $\mathbf{T_{\mathrm{N}}}$.} Semi-logarithmic plots of the magnetic contributions to (a-c) the thermal expansion coefficients \aimag, $i = a, b, c$, (d) the magnetic volume expansion coefficient $\beta_{\mathrm{mag}}$ and (e) the specific heat versus reduced temperature $|t| = |T/T_{\mathrm{N}}-1|$, for \TN\ $= 26.4$~K. Black squares (red circles) mark data for $T < T_{\mathrm{N}}$ ($T > T_{\mathrm{N}}$). Red and black solid lines are fits to the data according to Eq.~\eqref{eq:Scaling}, vertical dashed lines indicate the fitting window. Values for $\alpha^\pm$ with no error given were fixed for the fitting.}
\end{figure*}

The fits to the critical region around \TN, roughly around $0.01 < |t| < 0.1$, for the linear thermal expansion coefficients, the volume expansion coefficient and the specific heat are shown in Fig.~\ref{T_N_CritExp}.
A detailed overview over the best fit parameters and the fitting procedure is given in the supplement (Tab.~S3).
The two main results which stand out from the fits are that (1) the critical exponents $\alpha^\pm$ are all positive and lie between $\alpha^\pm = 0.088(7)$ and 0.23(4), and (2) the ratio of the amplitudes for $T > T_{\mathrm{N}}$ and $T < T_{\mathrm{N}}$, $A^{+}/A^{-}$, lies in the range from $A^{+}/A^{-} = 0.43(4)$ to 0.46(3) for the thermal expansion data, and $A^{+}/A^{-} = 0.34(1)$ for the specific heat. Considering these two main results, $\alpha^\pm$ values around 0.11 and ratios $A^{+}/A^{-}$ close to 0.5 -- but far away from 1 or even 1.5 -- enables us to clearly categorize the transition around \TN\ as a transition with a one-dimensional order parameter on a three-dimensional lattice, i.e., a 3D Ising-type transition.
This result is in line with the strong anisotropy visible in the magnetization and the phase diagrams of \fr\ as well as the strict (uniaxial) alignment of the Cu2 spins along the $c$ axis.

Previous calculations successfully described many of the properties of \ce{Cu3Bi(SeO3)O2X} (\ce{X}~=~\ce{Cl}, \ce{Br}) by an effective 3D spin model, i.e., by Heisenberg spins combined with Dzyaloshinskii-Moriya (DM) interactions~\cite{Rousochatzakis2015} and symmetric anisotropic exchange interactions~\cite{Constable2017}.
Also, neutron diffraction measurements suggested that the direction of the Cu spins is constrained to the $bc$ plane~\cite{Constable2017}.
For the brother compound \Brfran\ (CBSBr), on the other hand, a crossover from a 2D XY ($\beta_{\mathrm{c}} \leq 0.23$) to a 3D ($\beta_{\mathrm{c}} = 0.30$) character was suggested from neutron diffraction measurements near \TN~=~27.4~K~\cite{Pregelj2012}.
Deviations from the critical behavior are also observed in our data below \TN, roughly for $t < -0.1$. Together with the observation of dispersionless magnon modes along the $c$ direction at 2~K~\cite{Constable2017} this may point to 2D correlations evolving below \TN. However, whether or not the low-temperature behavior is caused by a 3D-to-2D crossover can not be concluded unambiguously from our data and necessitates further neutron or NMR studies.
Therefore, while a crossover to a spatial 2D behavior as reported for CBSBr cannot be neither concluded nor excluded from our data on \fr, our results clearly point to a 3D Ising-like behavior in the critical region around \TN.

\section{Conclusions}
The buckled-kagom\'{e} antiferromagnet \fran\ was investigated by thermal expansion, magnetostriction, magnetization and specific heat measurements. Its highly anisotropic lattice changes in temperature and in magnetic field reveal two phase transitions at \TN\ $= 26.4(3)$~K and \TS\ $= 120.7(5)$~K. The low-temperature and low-field AFM phase for $B\parallel c$, which experiences a field-driven metamagnetic transition to an FM phase, exhibits linear magnetoelastic coupling and a sizable mixed phase between the AFM and field-induced FM phases. Uniaxial pressure $p\parallel c$ stabilizes the AFM phase at the expense of the surrounding FM and paramagnetic (PM) phases. 
\TS\ is not affected by magnetic fields but strongly suppressed by uniaxial pressure along the in-plane directions, whereas $p\parallel c$ strongly enhances it by about 27~K/GPa.
The critical behavior in the vicinity of \TN\ is in line with calculations for the 3D Ising model.
Furthermore, while in-plane magnetic correlations extend to temperatures far above \TN\ the $c$ axis shows no significant magnetostriction and effects of a magnetic field on the thermal expansion above \TN.

The francisite \fran\ is an exciting compound with strong physical effects from both lattice and spin degrees of freedom at different temperatures.
Our study provides additional evidence that the low-temperature phase of \fran\ presents a versatile playground for investigating multiferroic effects, combining both a linear magnetoelastic and linear magnetoelectric coupling.

\section{Methods}
\subsection{Crystal growth}
Single crystals of \fran\ were grown by the chemical vapor transport method as reported in Ref.~\onlinecite{Gnezdilov2017}.
Different single crystals of sizes $2.30 \times 1.80 \times 0.58$~mm$^3$ ($m = 7.76(5)$~mg), $1.0 \times 1.0 \times 0.12$~mm$^3$ ($m = 0.60(5)$~mg) and $2.50 \times 2.00 \times 0.60$~mm$^3$ ($m = 8.88(5)$~mg) were used for our studies.

\subsection{High-resolution dilatometry}
High-resolution capacitance dilatometry measurements were performed in two three-terminal high-resolution capacitance dilatometers from Kuechler Innovative Measurement Technology with a home-built setup placed inside a variable temperature insert of an Oxford magnet system~\cite{Kuechler2017, Werner2017}. Linear thermal expansion coefficients \ali~$= 1/L_i \times  dL_i(T)/dT$ were derived for all crystallographic axes, i.e., $i = a, b, c$, in a temperature range from 2~K to 200~K, with magnetic fields up to 15~T applied along the measurement direction.
Field-induced length changes $dL_i(B)$ were measured at various fixed temperatures between 1.7~K and 200~K and the magnetostriction coefficients \li~$= 1/L_i \times dL_i(B_i)/dB_i$ were derived.

\subsection{Magnetization}
Magnetization measurements were performed in a Magnetic Properties Measurement System (MPMS3, Quantum Design) up to 7~T and in a Physical Property Measurement System (PPMS-14, Quantum Design) in fields up to 14~T.
A rotatable sample holder was used in the MPMS3 for measurements perpendicular to the $c$ axis in order to determine the $a$ and $b$ axis for further magnetization measurements. Laue XRD measurements confirmed the orientation and quality of the crystals (see supplement, Fig.~S1).

\subsection{Specific Heat}
Specific heat measurements were performed on a PPMS calorimeter using a relaxation method, on a sample of $m = 1.8$~mg.

\bibliography{FrancisiteBib_npj.bib}

\section{Data Availability}
All data used in this study are available from the corresponding author upon reasonable request.

\section{Acknowledgments}
We acknowledge financial support by BMBF via the project SpinFun (13XP5088) and by Deutsche Forschungsgemeinschaft (DFG) under Germany’s Excellence Strategy EXC2181/1-390900948 (the Heidelberg STRUCTURES Excellence Cluster) and through project KL 1824/13-1. For the publication fee we acknowledge financial support by DFG within the funding programme „Open Access Publikationskosten“ as well as by Heidelberg University. Support by the P220 program of the Government of Russia through the project 075-15-2021-604 is acknowledged. MM acknowledges support by the Russian Foundation for Basic Research through grant 20-02-00015. 

\section{Author Contributions}
S.Sp. performed dilatometry, Laue x-ray diffraction, and magnetization measurements together with their analysis. P.B., M.M., and A.V. synthesized and characterized the single crystals. M.M. performed specific heat measurements. S.Sp. wrote the draft. R.K. supervised the project. All authors reviewed and commented on the manuscript.

\section{Competing Interests Statement}
The authors declare no competing interests.

\clearpage

\end{document}


\title{Supplemental Material: Linear magnetoelastic coupling and magnetic phase diagrams of the buckled-kagom\'e antiferromagnet \ce{Cu3Bi(SeO3)2O2Cl}}
	\author{S. Spachmann$^{1,*}$}
	\author{P. Berdonosov$^{2,3}$}
	\author{M. Markina$^{2}$}
	\author{A. Vasiliev$^{2,3}$}
	\author{R. Klingeler$^{1}$}
	
	\affiliation{$^1$Kirchhoff Institute for Physics, Heidelberg University, D-69120 Heidelberg, Germany}
	\affiliation{$^2$Lomonosov Moscow State University, Moscow 119991, Russia}
	\affiliation{$^3$National University of Science and Technology "MISiS", Moscow 119049, Russia}
	\date{\today}

\maketitle
\beginsupplement

\section{Sample Orientation}
\begin{figure}[htbp]
	\center{\includegraphics [width= 0.95\columnwidth,clip]{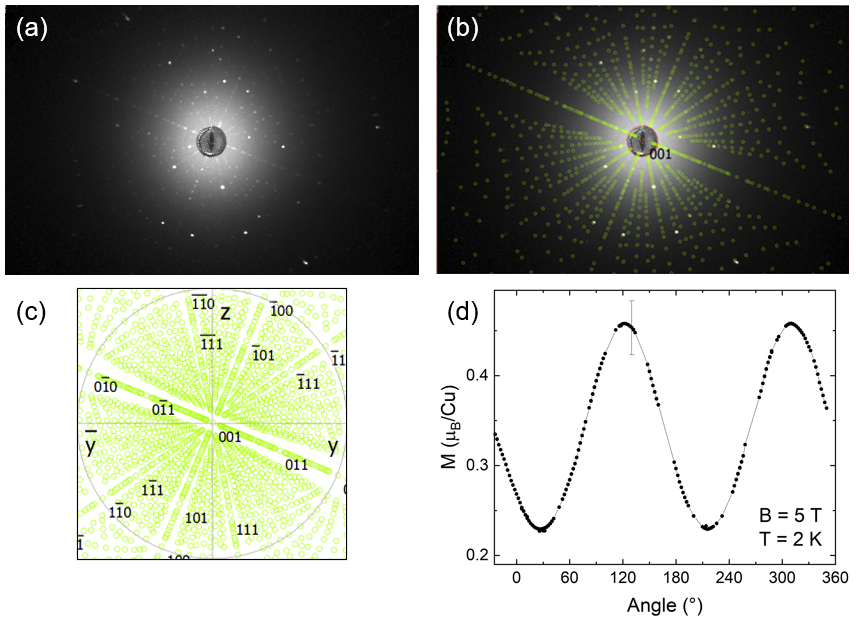}}
	\caption[] {\label{Laue-and-M-Angle} Orienting the samples: (a) Exemplary Laue XRD image and (b) corresponding orientation depicted in the Laue image itself and (c) a wider stereographic projection. (d) In-plane angular dependence of the magnetization at 2~K and 5~T. The error bar is representative for all data points and mainly results from mass determination.}
\end{figure}

Laue x-ray diffraction (XRD) and angle-dependent magnetization measurements at $T =2$~K and $B =5$~T are shown in Fig.~\ref{Laue-and-M-Angle}.
Laue XRD data were used to orient the samples precisely for in-plane thermal expansion along the $a$ and $b$ axis, whereas the angle-dependent magnetization (Fig.~\ref{Laue-and-M-Angle}(d)) served to orient the samples for magnetization measurements along the easy ($b$ axis) and hard ($a$ axis) in-plane directions.

\section{Lattice and Couplings}
\begin{figure}[htbp]
	\center{\includegraphics [width= 0.95\columnwidth,clip]{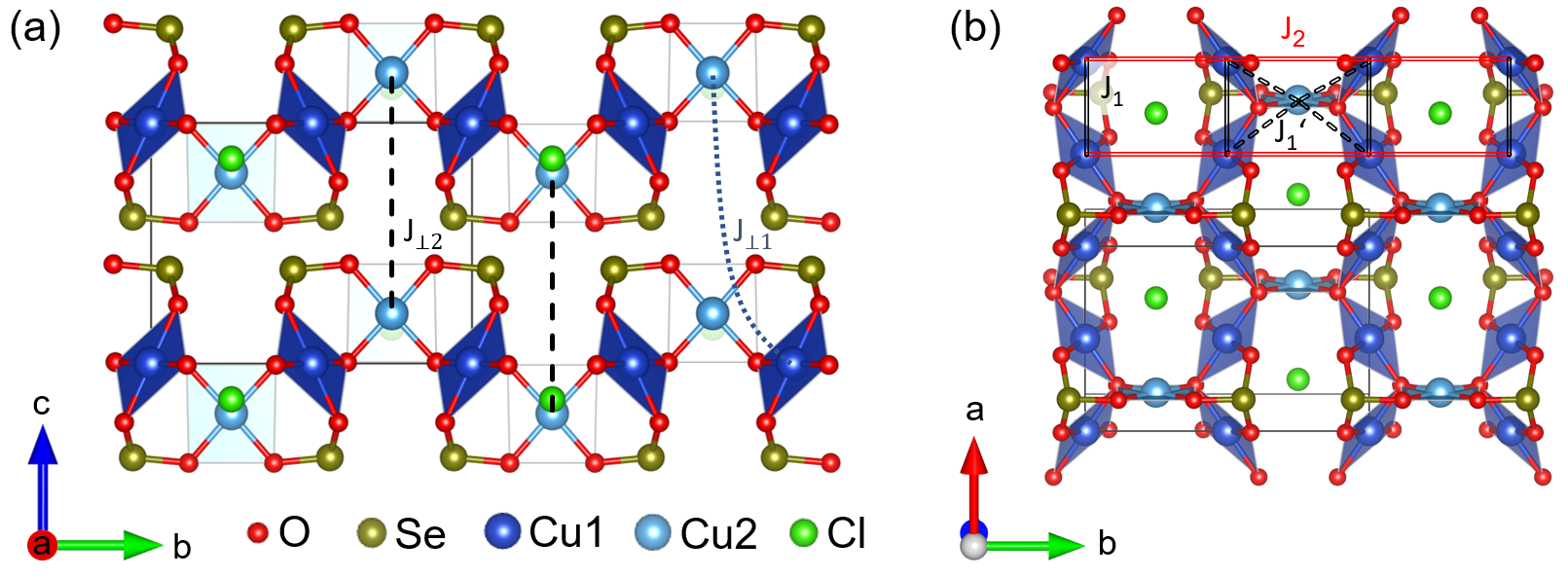}}
	\caption[] {\label{lattice_couplings} Crystal structure of \fran\ for (a) the $bc$-plane and (b) the $ab$-plane. Thin black lines mark the non-magnetic high temperature orthorhombic $Pmmn$ unit cell. Black, red and blue lines mark different exchange couplings. Bi ions are omitted for clarity and (b) is slightly tilted forward for better visibility. Visualization by VESTA.~\cite{VESTA}}
\end{figure}

The lattice of the francisite \fran\ according to Ref.~\onlinecite{Pring1990} along with the dominant couplings is presented in Fig.~\ref{lattice_couplings}.

\section{Additional Thermal Expansion Data}

\subsection{Volume Thermal Expansion}
\begin{figure}[htbp]
	\center{\includegraphics [width=0.5\columnwidth,clip]{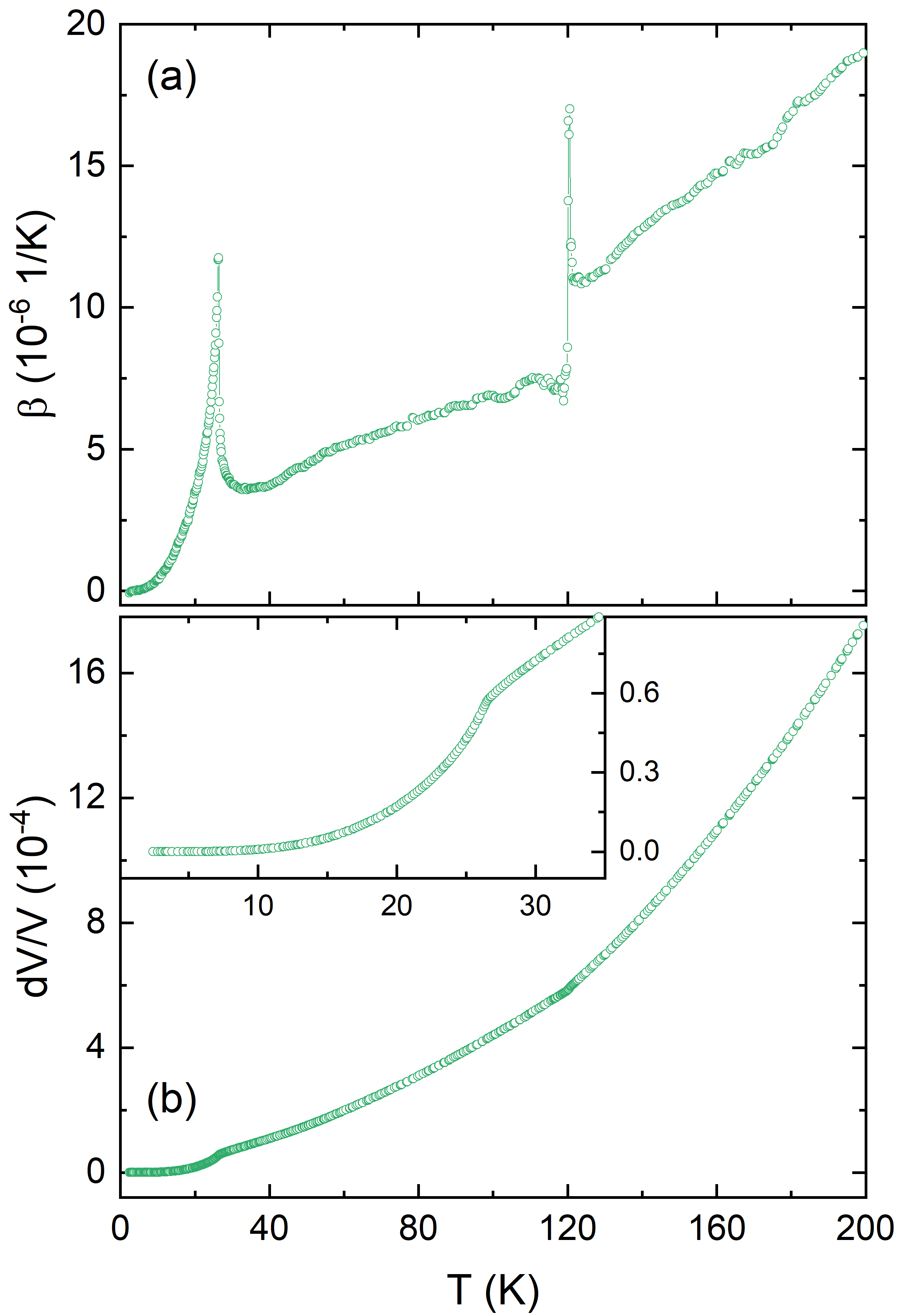}}
	\caption[] {\label{beta_0T} (a) Volume thermal expansion coefficient $\beta$ and (b) volume changes $dV/V$ in zero-field, calculated by adding \ali, i = $a$, $b$, $c$, and subsequent integration to obtain $dV/V$. The inset in (b) shows a magnification of the low temperature window.}
\end{figure}

The volume thermal expansion, gained from summing over the $a$, $b$, and $c$ axis, is shown in Fig.~\ref{beta_0T}. A strong kink is seen in $dV/V$ at \TN\ (Fig.~\ref{beta_0T}(b)), corresponding to a positive $\lambda$-like anomaly in the volume expansion coefficient $\beta$ (Fig.~\ref{beta_0T}(a)). At \TS\ a jump in $\beta$ occurs. The additional peak is most likely an artifact from summing the three linear thermal expansion coefficients of the $a$, $b$, and $c$ axis to obtain the volume expansion.

\subsection{Phononic Background Fits}
\begin{figure}[htbp]
    \center{\includegraphics [width=0.5\columnwidth,clip]{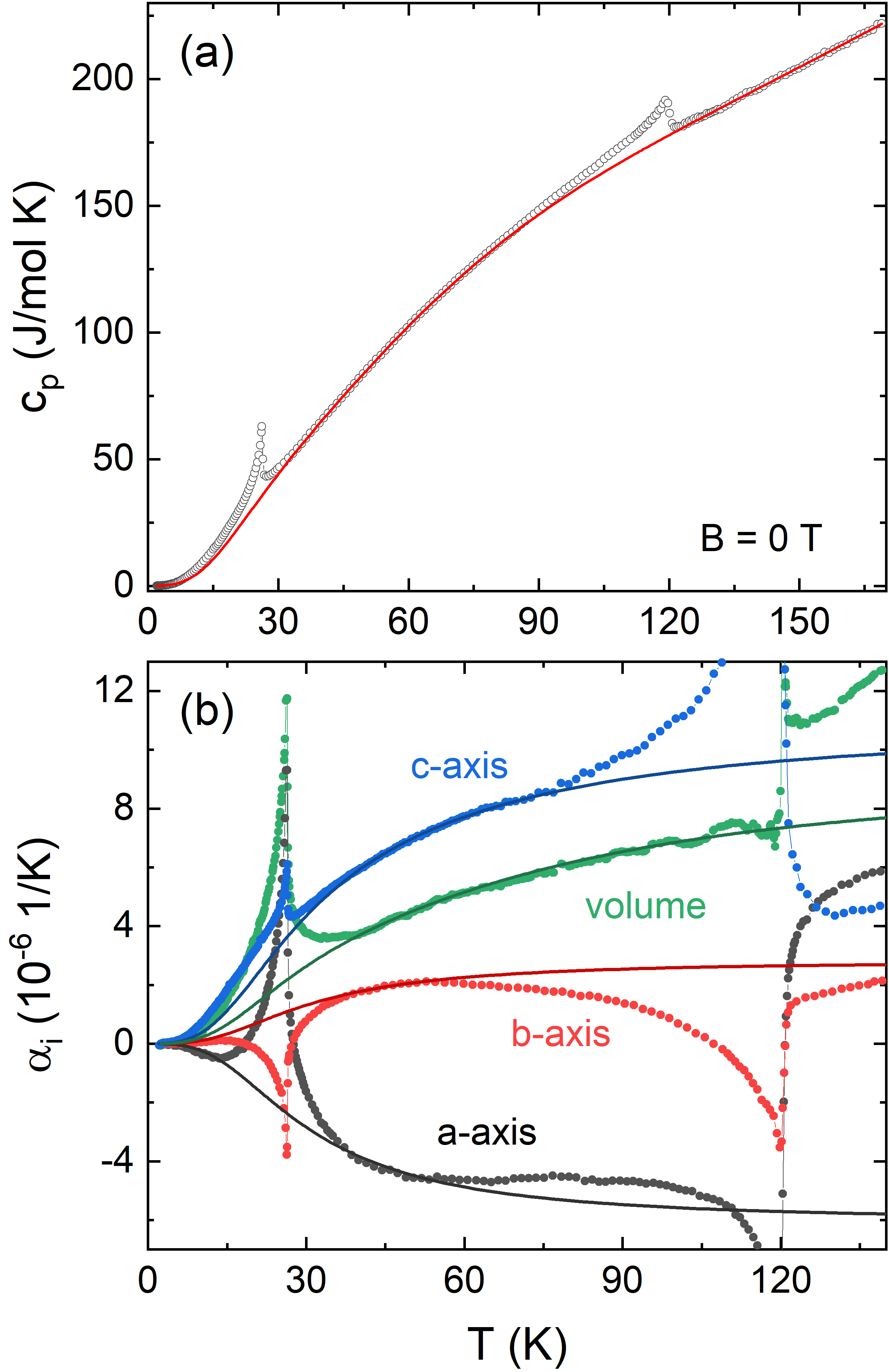}}
	\caption[] {\label{Alpha_c_p_bgs} Phononic background fits (lines) to (a) the specific heat and (b) the low temperature thermal expansion data (circles) in zero-field as explained in the text.}
\end{figure}

Phononic background fits to the thermal expansion coefficients and the specific heat are shown in Fig.~\ref{Alpha_c_p_bgs}.
Both thermal expansion and specific heat are fitted by Debye and Einstein contributions according to 
\begin{equation}
c_p^{ph}= n_{\mathrm{D1}}D\left(\frac{T}{\Theta_{D1}}\right) + n_{\mathrm{D2}}D\left(\frac{T}{\Theta_{D2}}\right)+n_{\mathrm{E}}E\left(\frac{T}{\Theta_{E}}\right),
\end{equation}
where $n_{\mathrm{D1,2}}$ and $n_{\mathrm{E}}$ are constants, and $D(T/\Theta_{\mathrm{D1,2}})$ and $E(T/\Theta_{\mathrm{E}})$ are the Debye and Einstein functions with the Debye and Einstein temperatures $\Theta_{\mathrm{D1,2}}$ and $\Theta_{\mathrm{E}}$.
A fit to the specific heat (Fig.~\ref{Alpha_c_p_bgs}(a)) below the onset of \TS, at 35~K to 93~K, yields n$_{\mathrm{D1}}$~=~3.25, n$_{\mathrm{D2}}$~=~6.14 and n$_{\mathrm{E}}$~=~12.1 with $\Theta_{\mathrm{D1}}$~=~127~K, $\Theta_{\mathrm{D2}}$~=~377~K and $\Theta_{\mathrm{E}}$~=~1063~K.
These Debye and Einstein temperatures are then used to fit the thermal expansion data in the range from 35~K to 60~K.
Due to the high value of the Einstein temperature, however, the contribution of Einstein modes to the thermal expansion below 100~K is negligible. Therefore, and in order to reduce the number of free parameters, the Einstein mode is omitted for the thermal expansion fit.
The resulting background fits are shown in Fig.~\ref{Alpha_c_p_bgs}(b).
The fits for the $c$ axis and volume describe the data very well -- the $c$ axis up to about 80~K and the volume up to \TS\ -- while the $a$ and $b$ axis only coincide with the data in a narrow temperature range, roughly between 40~K and 54~K.

\subsection{Extraction of Jump Heights}
\begin{figure}[htbp]
	\center{\includegraphics [width=0.7\columnwidth,clip]{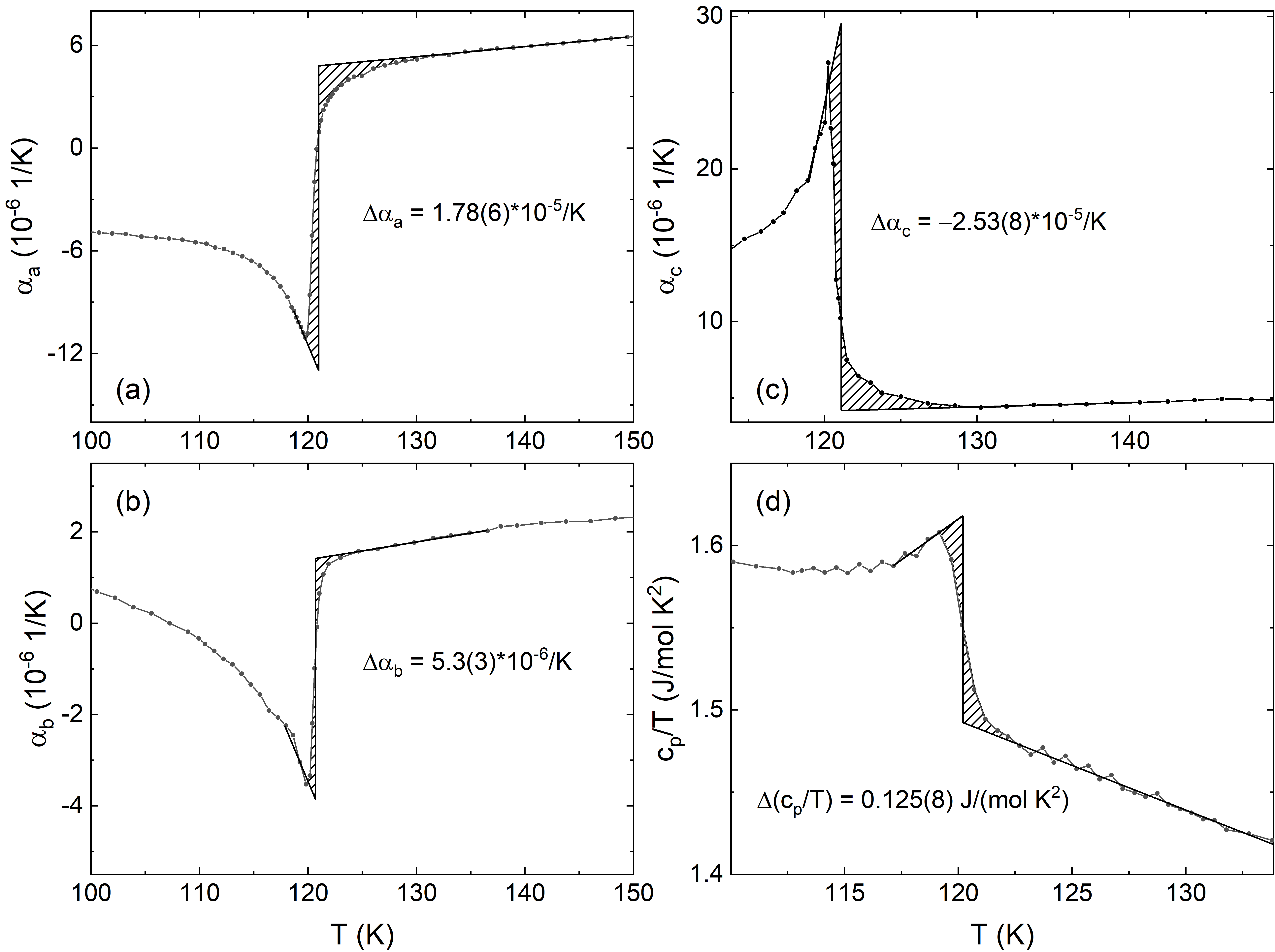}}
	\caption[] {\label{Jumps_Ts} Determination of the jump heights in  (a-c) thermal expansion and (d) specific heat by an area-conserving method.}
\end{figure}

The extraction of jump heights at \TS\ by an area-conserving (and entropy-conserving for \cp) method for the thermal expansion and specific heat is shown in Fig.~\ref{Jumps_Ts}. Jump heights are indicated in the figure.

\subsection{Gr\"{u}neisen Ratios}
\begin{figure}[htbp]
	\center{\includegraphics [width=0.4\columnwidth,clip]{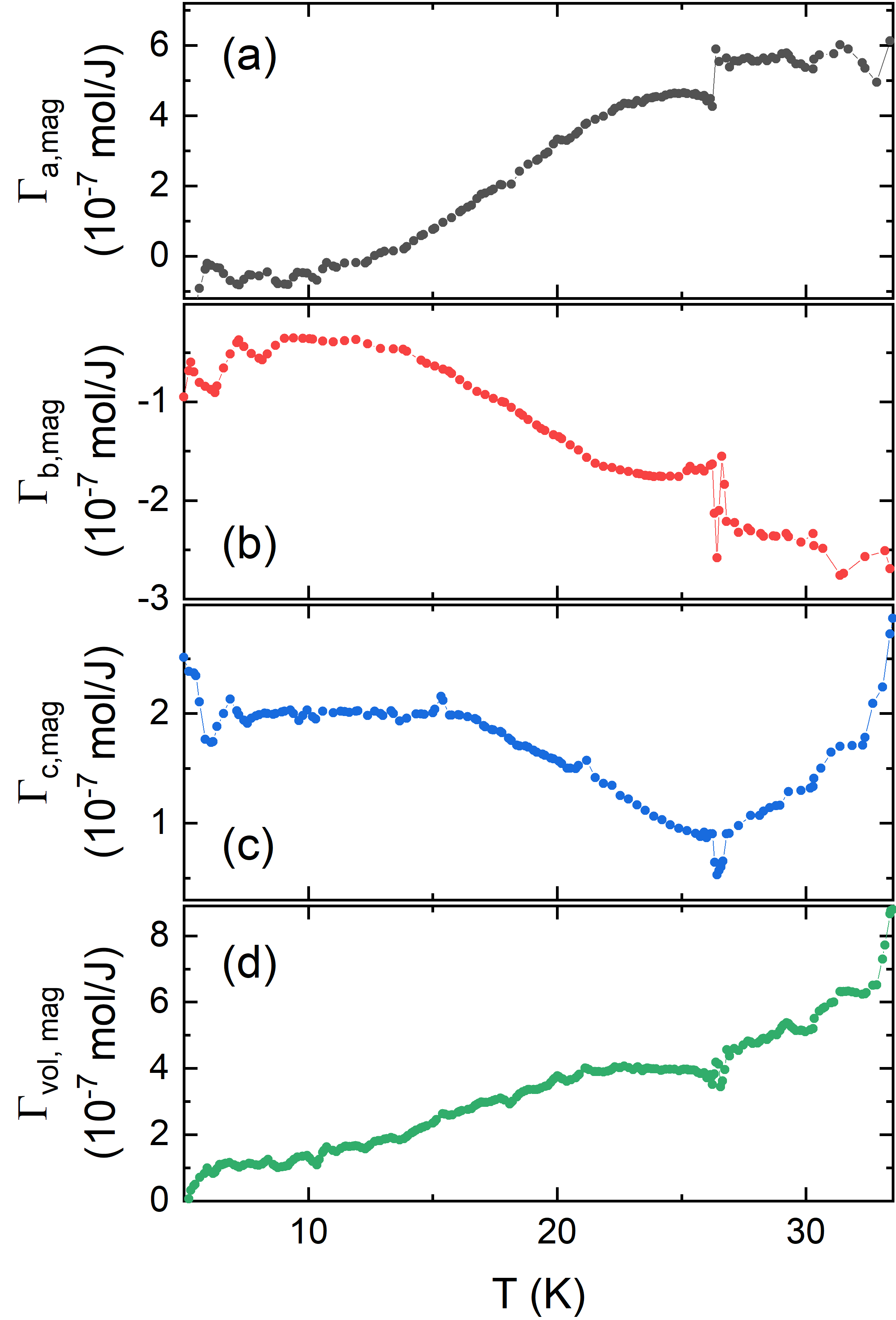}}
	\caption[] {\label{Gamma_mag} Magnetic Gr\"{u}neisen ratios $\Gamma_{\mathrm{mag},i} = \alpha_{\mathrm{mag},i}$/$c_{\mathrm{p,mag}}$ for (a) $a$, (b) $b$, and (c) $c$ axis, as well as for (d) the volume.}
\end{figure}

Fig.~\ref{Gamma_mag} shows the magnetic Gr\"{u}neisen ratios $\Gamma_{i,\mathrm{mag}} = \alpha_{i,\mathrm{mag}}/c_{\mathrm{p, mag}}$ for all axes and the volume. Notably, the $\Gamma_{i,\mathrm{mag}}$ reach a plateau around \TN\ (26.4~K) down to about 22~K, then vary upon cooling and reach another plateau below 15~K. The changes between 15~K and 22~K indicate the evolution of a competing energy scale in this regime.

\clearpage

\subsection{Thermal Expansion at 0~T and 15~T}
\begin{figure}[htbp]
	\center{\includegraphics [width=0.4\columnwidth,clip]{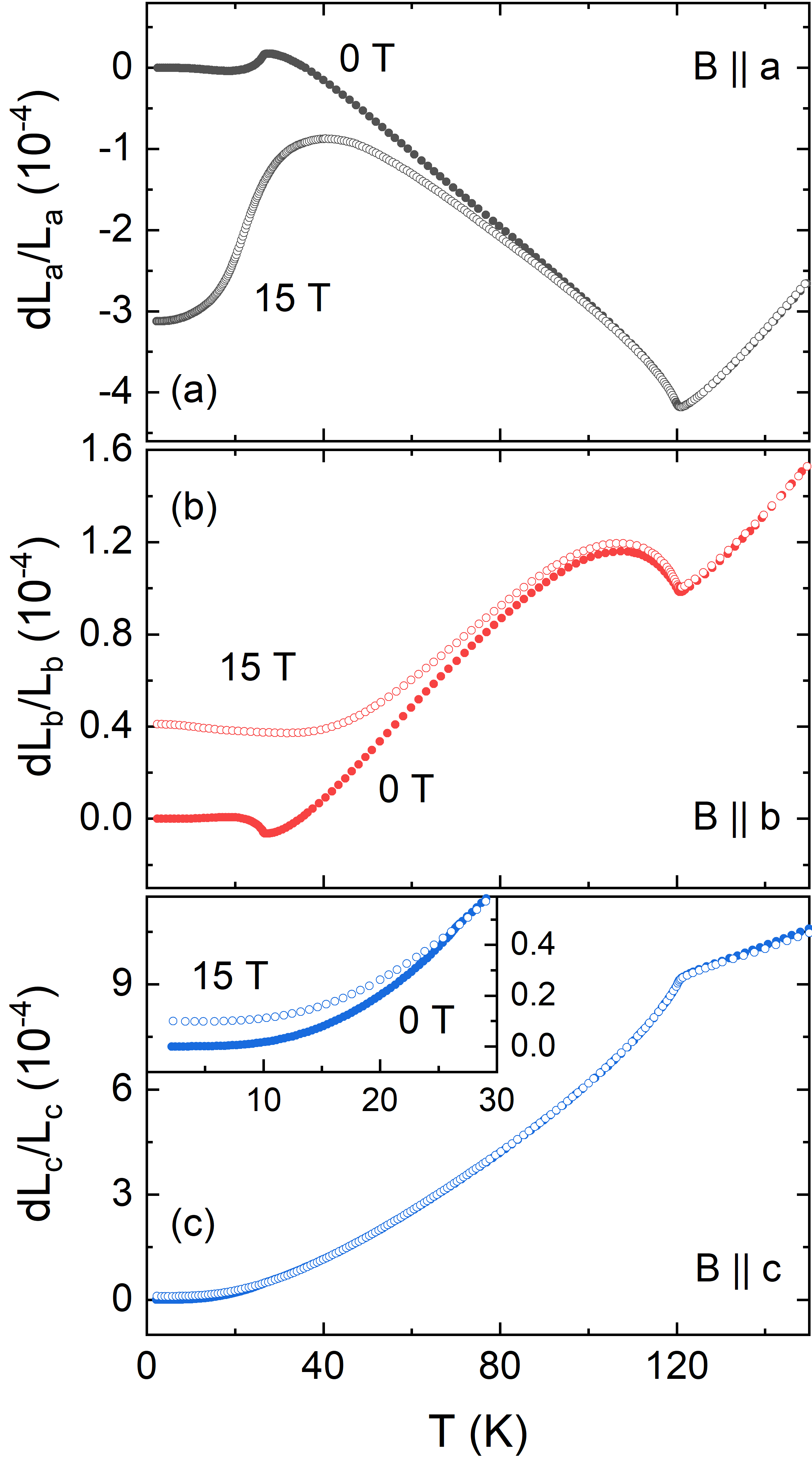}}
	\caption[] {\label{DLL_0T-15T} Comparison of relative length changes $dL_i/L_i$ in zero-field (closed circles) and B~=~15~T (open circles). Inset in (c) shows a magnification of the low temperature window.}
\end{figure}

A comparison of thermal expansion data at 0~T and 15~T, normalized above \TS, shows in-plane field effects up to about 100~K as described in the main article. (Fig.~\ref{DLL_0T-15T}).

\section{Failure of Grüneisen Scaling below $\mathbf{T_{\mathrm{N}}}$}
\begin{figure}[htbp]
	\center{\includegraphics [width=1\columnwidth,clip]{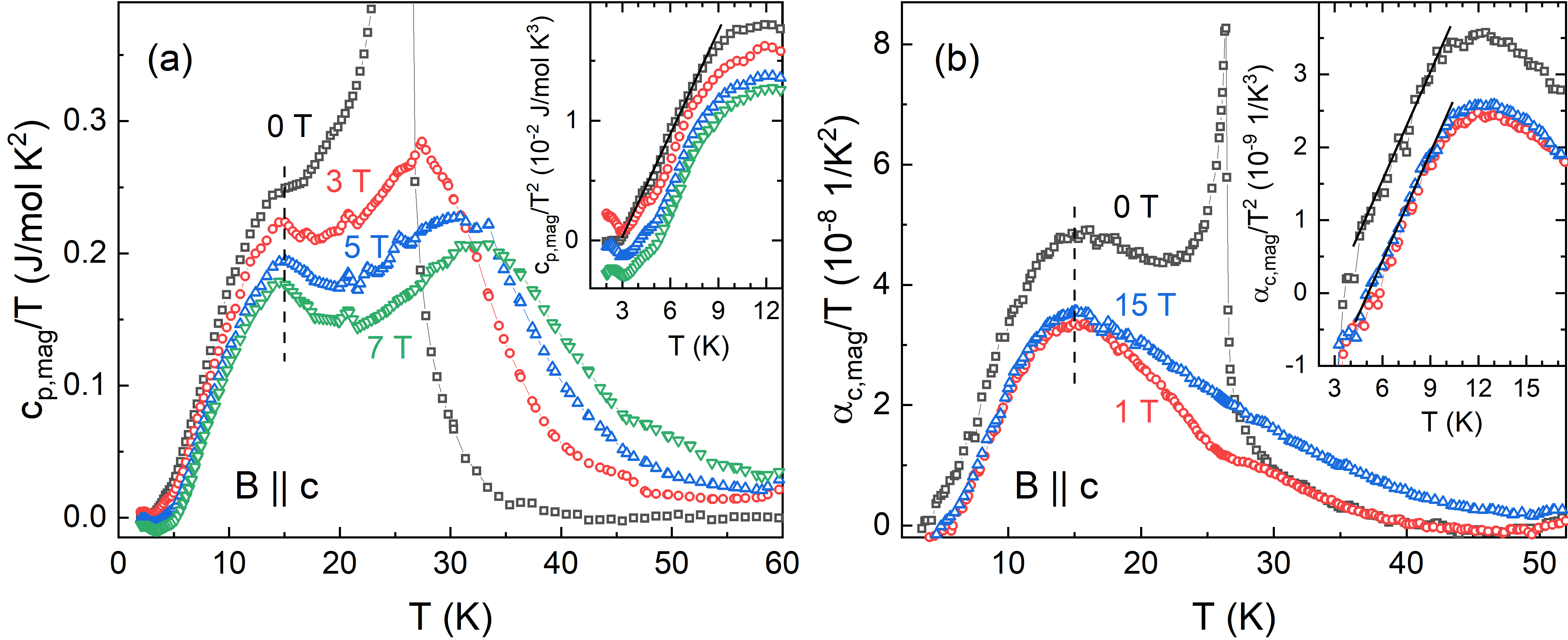}}
	\caption[] {\label{SI_Gruen_below_TN} Magnetic contributions to (a) the specific $c_{\mathrm{p}}/T$ and (b) the thermal expansion $\alpha_{\mathrm{c}}/T$ at zero-field and in fields $B\parallel c > 0$ as shown in the figure. Vertical dashed lines mark the peak position at 15~K. The insets show $c_{\mathrm{p,mag}}/T^{2}$ and $\alpha_{\mathrm{c,mag}}/T^{2}$ vs. $T$. Black lines are guides to the eye.}
\end{figure}

In the main text we mentioned the failure of Gr\"{u}neisen scaling below 22~K (see Fig.~2). Here, we want to spend some more time to explore this behavior.
Fig.~\ref{SI_Gruen_below_TN} shows the magnetic contributions to the specific heat and the thermal expansion in zero-field and at $B > 0$ up to 7~T and 15~T, respectively.
The same phononic background as described in the main text was used for all data sets.
For better comparability with $c_{\mathrm{p,mag}}/T$ we plotted $\alpha_{\mathrm{c,mag}}/T$ in Fig.~\ref{SI_Gruen_below_TN}(b).
The specific heat data shows two anomalies, one at 15~K which is not affected my the magnetic field and one which shifts to higher temperatures and broadens as the field is increased. The latter marks the crossover from the field-induced ferrimagnetic phase to the paramagnetic phase.
Similarly, $\alpha_{\mathrm{c,mag}}/T$ shows a peak at 15~K which is insensitive to a magnetic field and an additional shoulder may be seen around 30~K at 1~T, which is indistinguishable from the high-temperature tail of the 15~K peak at 15~T.
So what is the origin of the peak at 15~K?
A Schottky anomaly can be ruled out, because in contrast to the observed behavior it would shift in an applied magnetic field unless it were from transitions of electrons between energy levels of the same spin quantum number $m_S$.
Plotting $\alpha_{\mathrm{c,mag}}/T^{2}$ vs. $T$ (Fig.~\ref{SI_Gruen_below_TN}(b) inset) shows a linear rise up to about 10~K, i.e., $\alpha_{\mathrm{c,mag}} \propto T^{3}$, with a negative offset at $B > 0$. While the offset can be related to ferromagnetic magnons -- $\alpha_{\mathrm{FM}} \propto c_{\mathrm{p,FM}} \propto T^{3/2}$ -- the linear behavior signals either phononic or antiferromagnetic (AFM) magnonic contributions. In the ferromagnetic phase above the critical field no AFM magnons can be present, therefore the linear rise suggests a phononic origin of the 15~K anomaly judging from the thermal expansion data.
In contrast, the rise of $c_{\mathrm{p,mag}}/T^{2}$ (inset of Fig.~\ref{SI_Gruen_below_TN}(a)) is not as clearly linear and additional effects which do not show up in the thermal expansion may be present, leading to the failure of Gr\"{u}neisen scaling.

Considering results from other experimental techniques, there have been several observations reported in the AFM and FM phases for $B\parallel c$ in \fran\ which may be related to the anomaly:
Infrared reflection and transmission studies found two low-frequency phonon excitations at 5~K.~\cite{Miller2012} One of them, at 33.1~cm$^{-1}$ ($\approx 48$~K), is visible in zero-field and only shifts slightly, to about 35~cm$^{-1}$ in fields $B\parallel c$ of 10~T. For $B\perp c$ the resonance frequency of this phonon mode decreases. This phonon branch was tentatively assigned to magnon excitations~\cite{Miller2012}, based on an analysis of previously reported oscillator strengths of magnons and electromagnons. The second phonon excitation at 5~K was only observed for $B\parallel c$ in the ferromagnetic phase, with a frequency of 10.5~cm$^{-1}$ ($\approx 15.1$~K) at 1~T.~\cite{Miller2012} This phonon, however, shifts linearly in an applied magnetic field, up to about 16~cm$^{-1}$ (23~K) at 7~T.

Furthermore, a global spin gap of 1.57~meV (18.2~K) was reported from inelastic neutron scattering experiments.~\cite{Constable2017} This fits with the temperature scale of the anomaly and would suggest a relation to magnonic excitations.
The presence of the anomaly in zero-field, i.e., in the AFM phase, would then be explained by ferromagnetic spin-waves within the ferromagnetically coupled layers.

Lastly, two resonances at 1.23~meV (14.3~K) and 1.28~meV (14.8~K) were observed in the brother compound \Brfran\ in time-domain THz spectra at 3.9~K.~\cite{Wang2012}
Electron spin resonance (ESR) measurements from the same report suggest that these two resonances are of magnetic origin and shift to higher frequencies in higher magnetic fields.
However, a flat, i.e., field-independent, resonance would not be seen by spectroscopic field-sweeps around 300~GHz. Frequency sweeps at different magnetic fields would be necessary to observe such a resonance.

In conclusion, low-energy optical phonons, potentially coupled to magnon excitations, seem to be the cause for the anomalies observed around 15~K.

\section{Magnetization measurements}
\begin{figure}[htbp]
	\center{\includegraphics [width=0.9\columnwidth,clip]{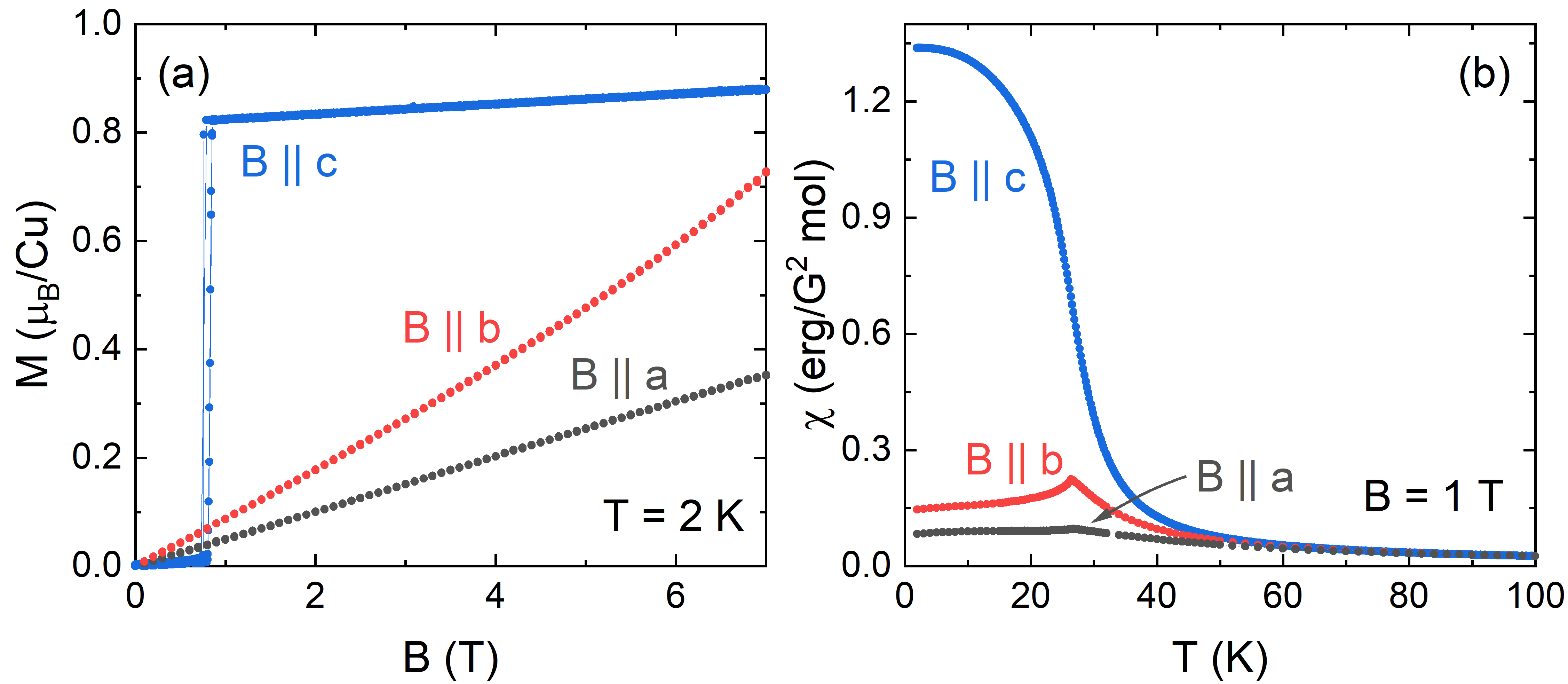}}
	\caption[] {\label{M-and-Chi_All} (a) Isothermal magnetization at T~=~2~K and (b) static magnetic susceptibility $\chi = M/H$ at $B~=~1$~T for $B\parallel a$ (black), $B\parallel b$ (red), and $B\parallel c$ (blue). Error bars are on the order of the size of the data points.}
\end{figure}

\begin{figure}[htbp]
	\center{\includegraphics [width=0.7\columnwidth,clip]{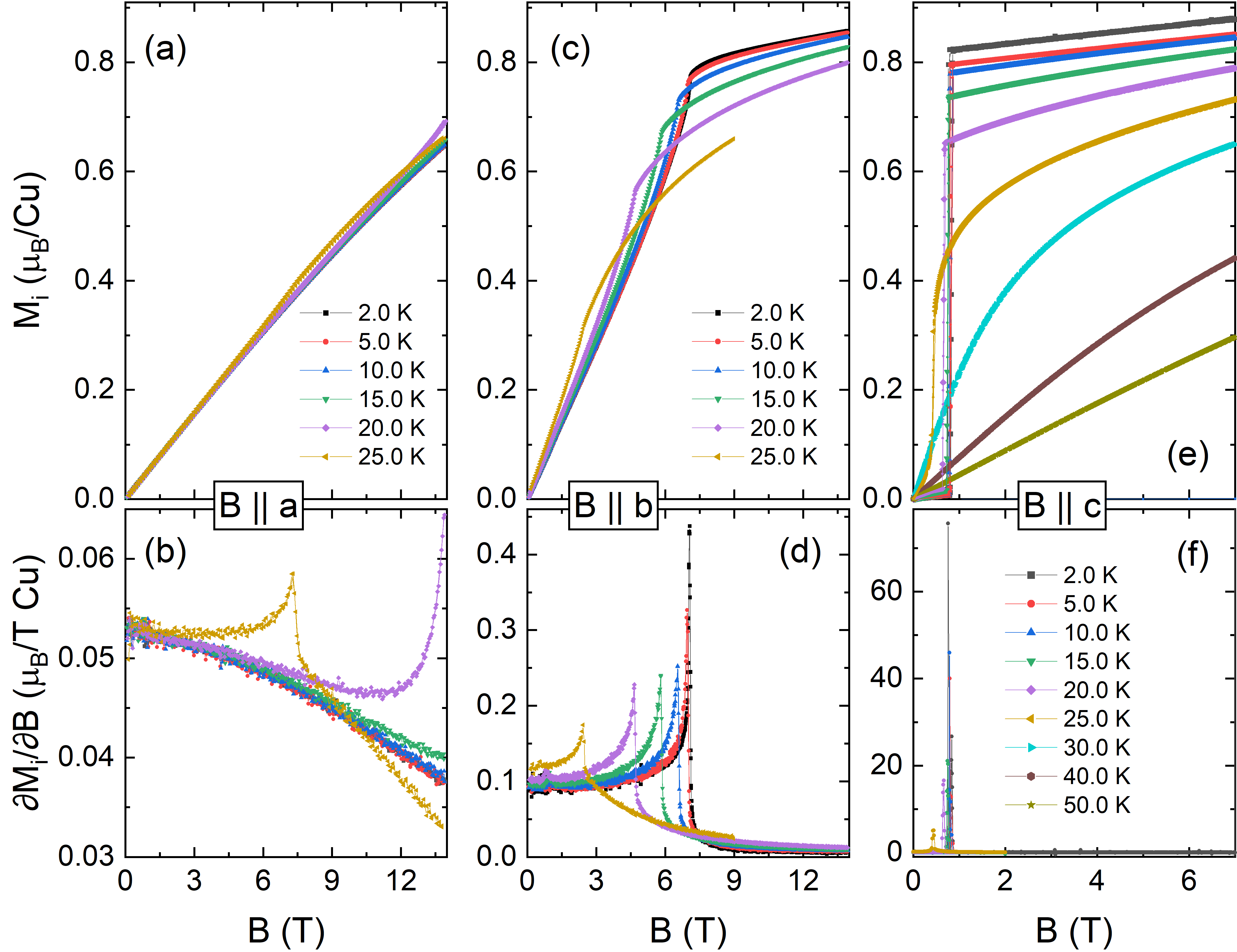}}
	\caption[] {\label{M-of-B_All} Isothermal magnetization $M_i$ and magnetic susceptibility ${\partial}M_i/{\partial}B$ at low temperatures for (a, b) $B\parallel a$, (c, d) $B\parallel b$ and (e, f) $B\parallel c$.}
\end{figure}

\begin{figure}[htbp]
	\center{\includegraphics [width=0.95\columnwidth,clip]{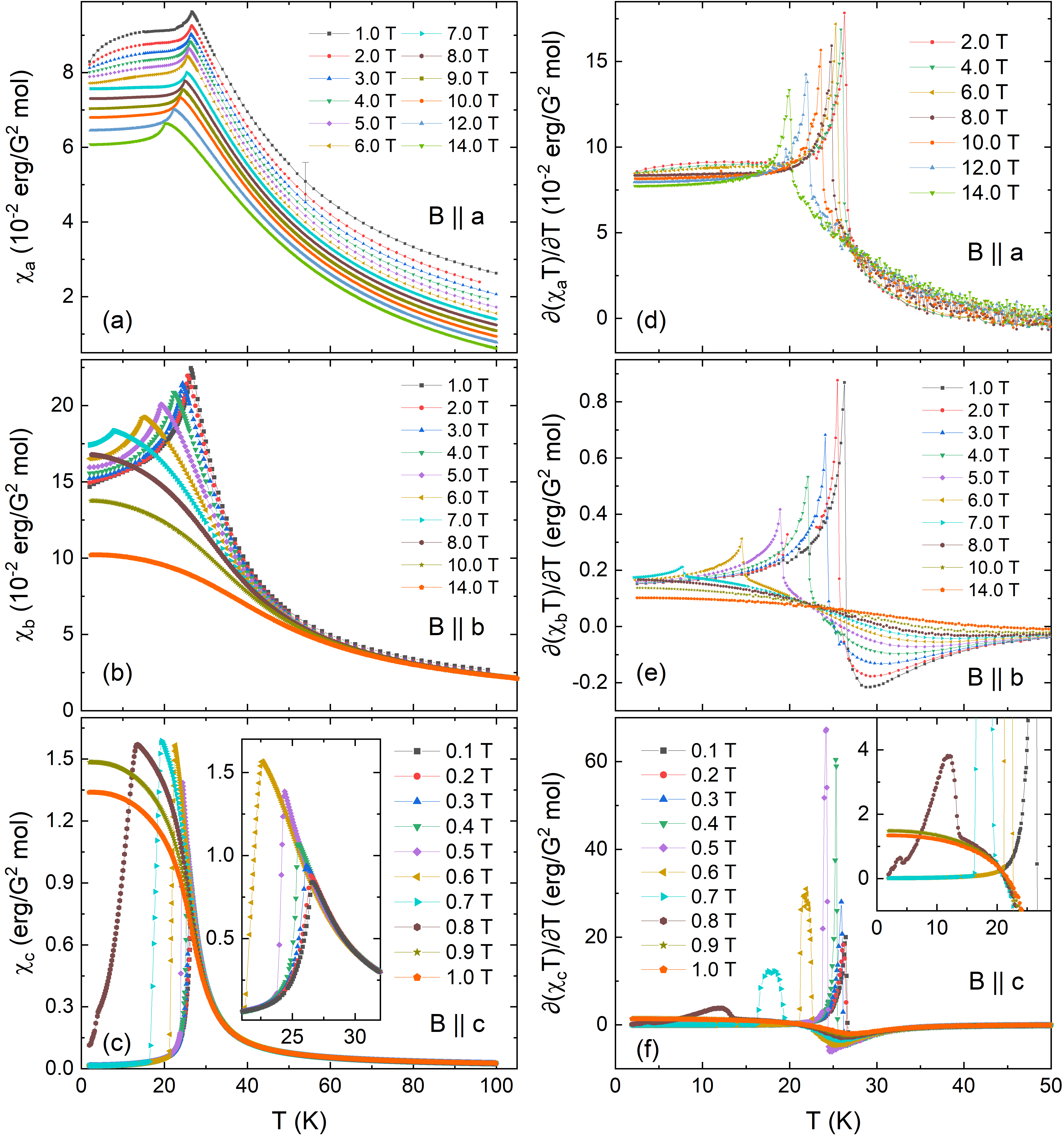}}
	\caption[] {\label{Chi_All} Static magnetic susceptibility $\chi = M/H$ and derived Fisher's specific heat in magnetic fields up to 7~T for $B\parallel a$ (a, d), $B\parallel b$ (b, e), and $B\parallel c$ (c, f). The inset in (c) shows a magnification for fields up to 0.6~T around \TN. The inset in (f) shows a magnification of the low temperature region. Data in (a) are offset by $-1.5\cdot 10^{-3}$~erg/(G$^2$ mol).}
\end{figure}

The isothermal magnetization at 2~K up to 7~T as well as the static magnetic susceptibility at 1~T are shown in Fig.~\ref{M-and-Chi_All}. 
Measurements of the isothermal magnetization up to 14~T at temperatures up to 25~K ($a$ and $b$ axis) and up to 50~K ($c$ axis) are shown in Fig.~\ref{M-of-B_All}.
Phase boundaries for the phase diagrams (Fig.~6) were obtained from the peaks in the magnetic susceptibility ${\partial}M_i/{\partial}B$.
The static magnetic susceptibility for all axes between 1~T and 14~T is shown together with the Fisher specific heat in Fig.~\ref{Chi_All}. Phase boundaries for the phase diagrams in Fig.~6 were obtained from the peaks in the Fisher specific heat for the in-plane directions and the $c$ axis up to 0.5~T. Above 0.5~T phase boundaries were obtained from the temperature at half of the jump heights upon entering and exiting the intermediate mixed phase.
Table~\ref{tab:fran_calculations} shows the experimental results from our magnetization measurements in comparison to the results of calculations by Nikolaev et al.~\cite{Nikolaev2016}

\renewcommand{\arraystretch}{1.2}
\begin{table*}[htb]
    \centering
     \caption{\label{tab:fran_calculations}Experimental results from our magnetization measurements in comparison to the results by Nikolaev et al.~\cite{Nikolaev2016}}
    \begin{tabular*}{0.6\textwidth}{l|@{\extracolsep{\fill}}cccc}
    \hline \hline
         & ${\partial}M_{\mathrm{a}}/{\partial}B$ & ${\partial}M_{\mathrm{b}}/{\partial}B$  & ${\partial}M_{\mathrm{c}}/{\partial}B$  & W$_{\mathrm{b}}$ \\
         & (\mB/T) & (\mB/T) & (\mB/T) & ({\mB}T) \\ \hline
        This work & 0.045(3) & 0.089(3) & 0.009(2) & 2.26 \\ 
        LDA+SO \cite{Nikolaev2016} & 0.15 & -  & 0.0073 & - \\ 
        Hartree Fock \cite{Nikolaev2016} & 0.044 & -  & 0.0061 & 2.2 \\ \hline
    \end{tabular*}
  
\end{table*}
\renewcommand{\arraystretch}{1}

\clearpage
\section{Quantitative analysis of the phase boundaries for $\mathbf{B\parallel c}$}

Table~\ref{tab:c-axis} shows the quantities extracted from magnetization and dilatometry measurements upon sweeping the magnetic field ($B$) or the temperature ($T$) and the resulting calculated quantities obtained according to Eq.~(2)~to~(4) in the main text.
\renewcommand{\arraystretch}{1.2}
\begin{table}[htbp]
	\centering
	\caption{\label{tab:c-axis} Jump heights, calculated changes in entropy, as well as field and pressure dependencies for the discontinuous transition in magnetic field from AFM to FM for $B\parallel c$ according to equations (2) to (4). 
	The quantity ${\partial}B_{\mathrm{crit}}/{\partial}T$ was calculated by taking the derivative of two polynomial fits in different temperature regimes to the values $B_{\mathrm{crit}}(T)$ in the phase diagram.}
	\begin{tabular*}{\textwidth}{c @{\extracolsep{\fill}} ccccccccc}
		\hline \hline
		\multicolumn{10}{c}{AFM $\rightarrow$ FM -- $c$ axis: Extracted and calculated quantities} \\
		T & B$_{\mathrm{crit}}$ & Sweep & ${\Delta}L_{\mathrm{c}}/L_{\mathrm{c}}$ & $\Delta M$ & ${\partial}B_{\mathrm{crit}}/{\partial}T$ & $\Delta S_{\mathrm{calc}}$ & ${\partial}T_{\mathrm{crit}}/{\partial}p_c$ & ${\partial}B_{\mathrm{crit}}/{\partial}p_c$ & ${\partial}$ln$(B_{\mathrm{crit}})/{\partial}p_c$ \\
		(K) & (T) &  & (10$^{-6}$) & ($\mu_{\mathrm{B}}$/Cu) & (mT/K) & (mJ/mol K) & (K/GPa) & (mT/GPa) & (\%/GPa) \\ \hline
		2.0$\pm$0.1 & 0.82 & $B$ & 7.8$\pm$0.5 & 0.82$\pm$0.02 & --2.2$\pm$1.0 & 30$\pm$14 & 34$\pm$16 & 76$\pm$6 & 9.3$\pm$0.7 \\ 
		5.0$\pm$0.1 & 0.81 & $B$ & 7.8$\pm$0.5 & 0.8$\pm$0.02 & --2.2$\pm$1.0 & 29$\pm$14  & 35$\pm$17 & 78$\pm$6 & 9.6$\pm$0.7 \\ 
		10.0$\pm$0.2 & 0.80 & $B$ & 7.5$\pm$0.5 & 0.78$\pm$0.02 & --7$\pm$3 & 90$\pm$40 & 11$\pm$5 & 77$\pm$6 & 9.6$\pm$0.7 \\ 
		15.0$\pm$0.2 & 0.75 & $B$ & 7.2$\pm$0.5 & 0.74$\pm$0.02 & --13$\pm$3 & 160$\pm$40 & 6.0$\pm$1.5 & 77$\pm$6 & 10.4$\pm$0.8\\ 
		17.9$\pm$0.4 & 0.70 & $T$ & 6.8$\pm$0.5 & 0.69$\pm$0.07 & --17$\pm$3 & 200$\pm$40 & 4.6$\pm$1.0 & 79$\pm$10 & 11.3$\pm$1.4 \\ 
		20.0$\pm$0.2 & 0.66 & $B$ & 5.8$\pm$0.4 & 0.65$\pm$0.02 & --27$\pm$8 & 290$\pm$90 & 2.6$\pm$0.8 & 71$\pm$6 & 10.8$\pm$0.9 \\
		21.9$\pm$0.3 & 0.60 & $T$ & 5.1$\pm$0.4 & 0.56$\pm$0.06 & --38$\pm$8 & 360$\pm$90 & 1.9$\pm$0.5 & 72$\pm$10 & 12.1$\pm$1.6 \\
		24.0$\pm$0.2 & 0.50 & $T$ & 3.1$\pm$0.4 & 0.36$\pm$0.04 & --60$\pm$20 & 360$\pm$130 & 1.1$\pm$0.5 & 69$\pm$12 & 14$\pm$3 \\ 
		25.0$\pm$0.2 & 0.44 & $B$ & 3.1$\pm$0.4 & 0.20$\pm$0.02 & --94$\pm$20 & 320$\pm$80 & 1.3$\pm$0.4 & 120$\pm$20 & 28$\pm$5 \\
		25.3$\pm$0.2 & 0.40 & $T$ & 1.2$\pm$0.3 & 0.12$\pm$0.04 & --142$\pm$30 & 280$\pm$120 & 0.6$\pm$0.3 & 80$\pm$40 & 20$\pm$9 \\ \hline
	\end{tabular*}

\end{table}
\renewcommand{\arraystretch}{1}

\clearpage
\section{Critical Scaling Analysis}

The magnetic contributions to the thermal expansion coefficients and the specific heat in Fig.~7 in the main text were fitted by Eq.~(5), i.e.,
\begin{equation*}
    c_{\mathrm{p}} = \frac{A^\pm}{\alpha^\pm}|t|^{-\alpha^\pm}(1+E^{\pm}|t|^{0.5})+B+D^{\pm}t
\end{equation*}
where $t = T/T_{\mathrm{N}}-1$ is the reduced temperature.
Initially, this left us with five free fit parameters for each fit.
However, since the phononic contributions to \cp\ and \ali\ were already subtracted, both the offset $B$ and the linear term $D^{\pm}$ were set to zero, leaving us with only three fit parameters.
To further reduce the free parameters to two, we tried fixing $\alpha^\pm$, either to values from fits for $T < T_{\mathrm{N}}$ or to a value giving good fit results over a wide range, also beyond the actual fitting range.
Our best least-square fitting results are presented in Tab.~\ref{tab:CritScaling}.
For comparison results with fixed and free values for $\alpha^\pm$ are shown for $\alpha_{\mathrm{a,mag}}$ and \cpmag.

\renewcommand{\arraystretch}{1.2}
\begin{table}[ht!]
    \centering
	\caption{\label{tab:CritScaling} Fit parameters for the critical scaling according to Eq.~(5) in the main text. Unless indicated otherwise, $B = 0$ and $D = 0$ were fixed for all fits. The parameters $A^\pm$ are in units of 1/K and J/(mol K), for thermal expansion and specific heat, respectively. 'Range' gives the fitting range. '(f)' indicates that the quantity was fixed manually and not fitted.}
    \begin{tabular}{lccccccccc}
    \hline \hline
         & Range $T < T_{\mathrm{N}}$ & $\alpha^{-}$ & $A^{-}$ & $E^{-}$ & Range $T > T_{\mathrm{N}}$ & $\alpha^{+}$ & $A^{+}$ & $E^{+}$ & $A^{+}/A^{-}$ \\ \hline
        $\alpha_{\mathrm{a,mag}}$ & $0.01 < |t| < 0.1$ & 0.088(7) & $7.7(4)\cdot 10^{-7}$ & $-1.77(3)$ & $0.01 < |t| < 0.1$ & 0.21(3) & $4.15(4)\cdot 10^{-7}$ & $-1.57(10)$ & 0.54 \\
         &  &  &  &  & $0.01 < |t| < 0.1$ & 0.088 (f) & $3.04(5)\cdot 10^{-7}$ & $-2.01(5)$ & 0.39 \\
         &  &  &  &  & $0.01 < |t| < 0.1$ & 0.116 (f) & $3.54(4)\cdot 10^{-7}$ & $-1.92(4)$ & 0.46 \\
        $\alpha_{\mathrm{b,mag}}$ & $0.08 < |t| < 0.1$ & 0.116 (f) & $-3.20(5)\cdot 10^{-7}$ & $-1.54(5)$ & $0.08 < |t| < 0.1$ & 0.11(2) & $-1.38(11)\cdot 10^{-7}$ & $-1.82(11)$ & 0.43 \\
        $\alpha_{\mathrm{c,mag}}$ & $0.006 < |t| < 0.06$ & 0.12(5) & $1.55(15)\cdot 10^{-7}$ & $-1.1(5)$ & $0.009 < |t| < 0.11$ & 0.23(4) & $6.7(3)\cdot 10^{-8}$ & $-0.9(3)$ & 0.43 \\
        $\beta_{\mathrm{mag}}$ & $0.01 < |t| < 0.13$ & 0.11(1) & $7.2(3)\cdot 10^{-7}$ & $-1.59(3)$ & $0.01 < |t| < 0.13$ & 0.20(2) & $3.24(4)\cdot 10^{-7}$ & $-1.38(9)$ & 0.45 \\
        $c_{\mathrm{p,mag}}$ & $0.006 < |t| < 0.1$ & 0.162(2) & 2.214(5) & $-1.49(2)$ & $0.01 < |t| < 0.15$ & 0.21(2) & 0.743(7) & $-1.60(4)$ & 0.34 \\
         & $0.006 < |t| < 0.1$ & 0.125 (f) & 2.16(2)$^a$ & $-2.06(5)$ & $0.01 < |t| < 0.15$ & 0.125 (f) & 0.663(6)$^b$ & $-2.06$ (f) & 0.31 \\ \hline
        \end{tabular}
        \begin{flushleft}
        \footnotesize{$^a$ Also fitting $D^{-}$, with $D^{-} = 26(4)$~J/(mol K)} \\
        \footnotesize{$^b$ Also fitting $D^{+}$, with $D^{+} = 4.2(3)$~J/(mol K)}
        \end{flushleft} 
\end{table}
\renewcommand{\arraystretch}{1}

\bibliographystyle{apsrev}
\bibliography{FrancisiteBib_npj.bib}